\documentclass[fleqn,usenatbib]{mnras}

\usepackage{newtxtext,newtxmath}
\usepackage{mwe}
\usepackage{subfig}
\usepackage[T1]{fontenc}
\usepackage{graphicx}
\DeclareRobustCommand{\VAN}[3]{#2}
\let\VANthebibliography\thebibliography
\def\thebibliography{\DeclareRobustCommand{\VAN}[3]{##3}\VANthebibliography}

\newcommand\fhr{\mbox{$\hspace{0.1cm} \!\!^{\mathrm h}$}}%
\newcommand\fmin{\mbox{$\hspace{0.1cm}\!\!^{\mathrm m}$}}%
\usepackage{graphicx}	
\usepackage{amsmath}	
\usepackage{threeparttable}
\usepackage{placeins}
\usepackage{graphicx}
\usepackage{wrapfig}
\usepackage{float}
\usepackage{gensymb}
\usepackage[dvipsnames]{xcolor}

\title[Radio emission from a post-merger galaxy]
{ Multifrequency analysis of the radio emission from a post-merger galaxy CGCG 292-057}

\author[Misra, Jamrozy and We\.zgowiec]{
Arpita Misra,$^{1}$\thanks{E-mail: arpita.misra@doctoral.uj.edu.pl}
Marek Jamrozy,$^{1}$
Marek We\.zgowiec$^{1}$
\\
$^{1}$Astronomical Observatory, Jagiellonian University, Orla 171, PL-30-244 Krakow, Poland \\
}

\date{Accepted XXX. Received YYY; in original form ZZZ}

\pubyear{2023}

\begin{document}
\label{firstpage}
\pagerange{\pageref{firstpage}--\pageref{lastpage}}
\maketitle

\begin{abstract}
 
Galaxies exhibiting a specific large-scale extended radio emission, such as X-shaped radio galaxies, belong to a rare class of winged radio galaxies. The morphological evolution of these radio sources is explained using several theoretical models, including galaxy mergers. However, such a direct link between a perturbed radio morphology and a galaxy merger remains observationally sparse. Here we investigate a unique radio galaxy J1159+5820, whose host CGCG 292-057 displays the optical signature of a post-merger system with a distinct tidal tail feature, and an X-shaped radio morphology accompanied by an additional pair of inner lobes. We observed the target on a wide range of radio frequencies ranging from 147 MHz to 4959 MHz, using dedicated GMRT and VLA observations, and supplemented it with publicly available survey data for broadband radio analysis. Particle injection models were fitted to radio spectra of lobes and different parts of the wings. Spectral ageing analysis performed on the lobes and the wings favors a fast jet realignment model with a reorientation timescale of a few million years. We present our results and discuss the possible mechanisms for the formation of the radio morphology.
 
 \end{abstract}




\section{Introduction}

Highly collimated large-scale  radio galaxy (RG) jets are formed from the relativistic outflows of charged particles and magnetic fields from the synchrotron emitting plasma of the active galactic nucleus (AGN). These jets ranging from a few kpc to Mpc in size are known to strikingly maintain their jet axis direction for almost $10^8$ years {(\citealt{2009MNRAS.395..812M}; \citealt{2021ApJS..255...22M}), or even $10^9$ years, as in the case of some giant RGs (\citealt{2015MNRAS.447.2468H}; \citealt{2019A&A...628A..69S})}. The typical hosts for RGs are giant ellipticals, and the most common mechanism for growing massive galaxies is galaxy mergers. Mergers invigorate galaxies with a fresh supply of gas and dust and can trigger AGN activity around the supermassive black hole (SMBH), ultimately playing a key r\^ole in explaining the growth of the SMBH at the center of most active galaxies (\citealt{2012MNRAS.426..276B}; \citealt{2013MNRAS.431.2661C}; \citealt{2019MNRAS.487.2491E}).
 A subset of such interactions can also take place in galaxies containing AGNs with radio jets, leading to non-standard radio morphologies as seen in the case of e.g. 3C 321 (\citealt{2008ApJ...675.1057E}), an FRII (\citealt{1974MNRAS.167P..31F}) type RG where one of the jets undergoes bending after interacting with a nearby companion galaxy going through a merger with its host. A similar but more severe bending of the jets is observed in the case of 3C 433 (\citealt{1983AJ.....88...40V}), which is a part of an interacting galaxy pair. During the merger process, the central AGN is bound to undergo disturbances and perturbations, and studying radio jets associated with such galaxies can act as an excellent tracer for understanding the central SMBH behavior.

\begin{figure*}
  \centering
  \includegraphics[scale=1.13]{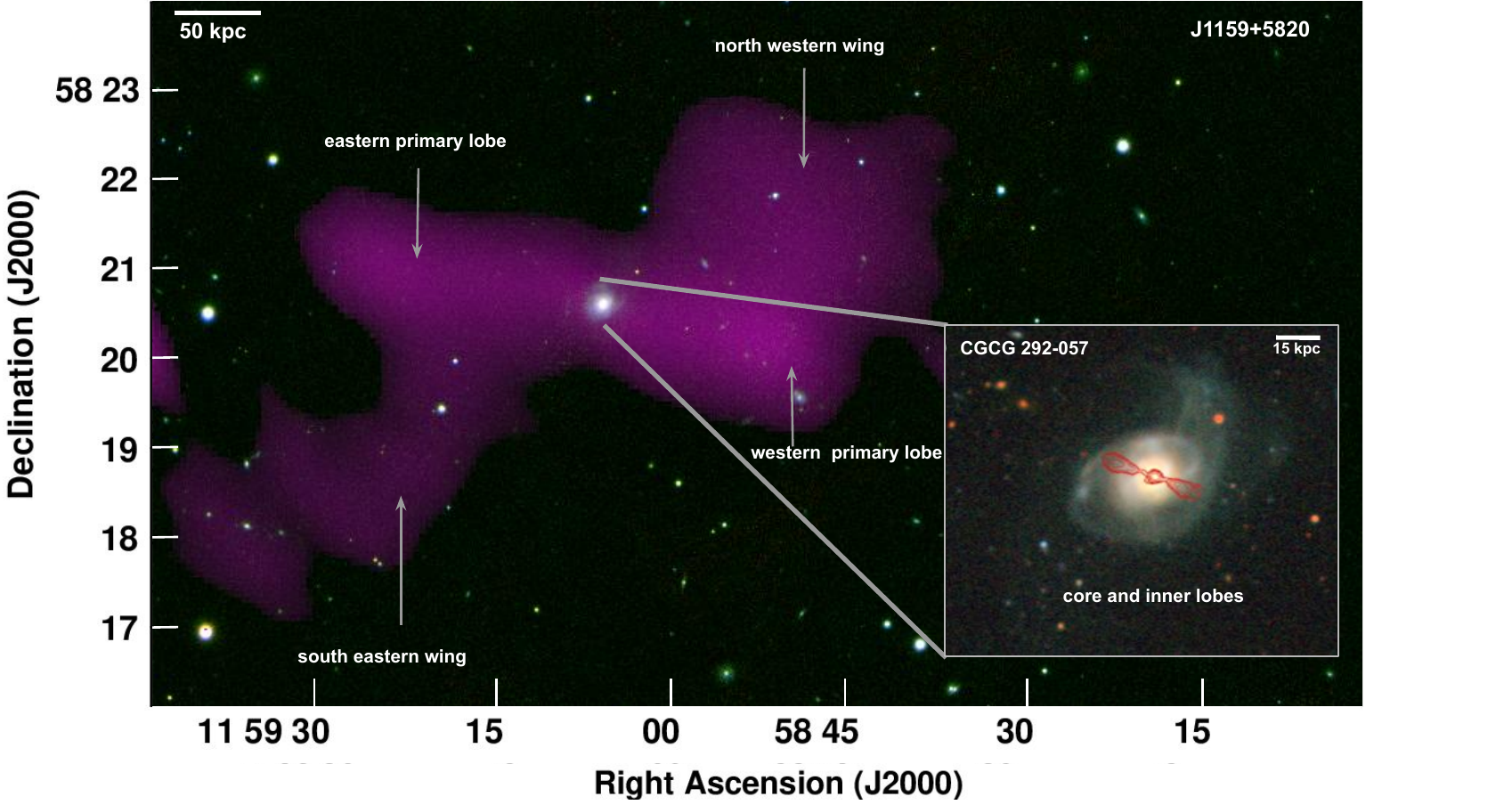}
  \caption{147 MHz GMRT radio map of J1159+5820 overlaid on PanSTARRS (\citealt{2016arXiv161205560C}) optical image. The right panel insert shows VLA contours at 4959 MHz of inner lobes and core (red contours) overlayed on the host galaxy CGCG 292-057 optical image from BASS (\citealt{2019ApJS..245....4Z}) with distinct tidal tail and shell features. }   
  \label{fig1}
\end{figure*}

The discovery of X-shaped radio sources (\citealt{1990Ap&SS.164..131F};  \citealt{1992ersf.meet..307L}; \citealt{2002A&A...394...39C}) poses a challenge towards understanding the dynamic interplay between radio jets, the central active region, and the intergalactic medium (IGM). The class of X-shaped RGs mainly consists of two misaligned pairs of lobes, the second pair having a relatively different orientation than the primary lobe, with the typical examples being 3C 223.1 and 3C 403 (\citealt{2002MNRAS.330..609D}). There are a few models that account for the formation of such sources, including (i) hydrodynamical backflow, where the secondary pair of lobes are formed by the backflow of the primary lobe that is deflected by the hot gas halo of the host galaxy (\citealt{1984MNRAS.210..929L}; \citealt{1995ApJ...449...93W}; \citealt{2020MNRAS.495.1271C}); (ii) spin flip of the primary black hole and quick realignment in a post-merger system (\citealt{2002PhDT.......178R}; \citealt{2002Sci...297.1310M}); (iii) presence of a dual AGN with two independent pairs of jets (  \citealt{2005MNRAS.356..232L}; \citealt{2007AJ....133.2097C}; \citealt{2022ApJ...933...98Y}); (iv) the projection due to a precessing beam (\citealt{1978Natur.276..588E}; \citealt{1985A&AS...59..511P}); and (v) the jet-shell interaction in a merged galaxy (\citealt{2012RAA....12..127G}). The list of such objects has so far constituted a small fraction of radio sources (\citealt{2007AJ....133.2097C}; \citealt{2012A&A...544A..36M};  \citealt{2018ApJ...852...48S};  \citealt{2019ApJS..245...17Y}) but with new highly resolved deep-sky surveys, X-shaped sources are found at an ever-increasing rate. 
 
 RGs usually go through an initial phase of jet activity, lasting between $\sim$10$^7$ and {$\sim$10$^9$ years ({\citealt{1986MNRAS.219..575C}}; \citealt{1999A&A...344....7P};  \citealt{2007A&A...470..875P})}, followed by a period of quiescence. However, in some RGs, nuclear activity can be classified as intermittent, where a new pair of radio lobes is visible along with the older pair, tracing plasma from a previous episode of jet activity.  This suggests a short dormant phase followed by restarting nuclear activity, leading to a double-double radio morphology (DDRG;     \citealt{2000MNRAS.315..371S}; \citealt{2009BASI...37...63S}; \citealt{2017MNRAS.471.3806K}; \citealt{2019A&A...622A..13M}). Identifying and studying the radio spectra of such galaxies is of great importance for understanding the duty cycle of RGs.

In this paper, we examine radio data of CGCG\,292-057, host of  J1159+5820, which is a unique RG that presents a non-standard X-shaped radio morphology and also shows a restarting pair of radio lobes as a DDRG. The target source displays  low-surface-brightness radio wings that span $\sim$7$'$ and high-surface-brightness lobes that span to  5$'$.  First described in the paper of \cite{2012MNRAS.422.1546K}, this object at RA = $11\fhr59\fmin05\fs7$, Dec = $+$58\degr20\arcmin36\arcsec (J2000.0) with z = 0.0537 (\citealt{2008ApJS..175..297A}) is classified as a low ionization nuclear emission line region (LINER) galaxy, based on its emission line fluxes.
 The host galaxy morphology was determined by the concentration index of about 2.81±0.03, which places it almost exactly at the separation line between elliptical and spiral galaxies. In the optical observations performed by \cite{2012MNRAS.422.1546K}, a face-on galaxy with a clear tidal tail feature was revealed, indicating traces of a merging event in the past. \cite{2018MNRAS.475.5179S}
found an offset of 291 mas in the optical position of the CGCG\,292-057 galaxy center given by the SDSS catalog and  8.46 GHz radio position of core given by the Cosmic-Lens All Sky Survey (CLASS; \citealt{2003MNRAS.341....1M}; \citealt{2003MNRAS.341...13B}) catalog.  Taking into account the velocity dispersion of 242±13.6 km/s, the mass of the central black hole was calculated as log(\( M_{BH}/M_\odot\)) = 8.47±0.32 (\citealt{2012MNRAS.422.1546K}). The target is also fairly isolated as it is not associated with any nearby group or cluster. The star formation rate was calculated by \cite{2015MNRAS.454.1556S}, with the use of mid-IR and UV emission, to be 2.89 and 1.50 \(M_\odot\) per year respectively, with evolved stellar composition at the center of the host galaxy.  X-ray studies conducted by \cite{2020ApJ...905..148B} detected an excess of X-ray emission in the bulge from a thermal plasma and the presence of a Fe XXV K$\alpha$ line. From the SDSS spectrum, doubly peaked emission lines were observed which might be considered as an indicator for an SMBH binary (\citealt{2012MNRAS.422.1546K}).

In this paper, we present a multifrequency radio analysis of J1159+5820, which exhibits a rare combination of DDRG along with an X-shaped morphology. The presence of all of these features makes the target an ideal source for studying X-shaped RGs connected with galaxy mergers. We used archival and dedicated radio data at a range of frequencies between 54 MHz and 8440 MHz, revealing its compact (a few kpc) and extended (several hundred kpc) radio structure. By analyzing its radio spectra, we estimate the ageing of the electron population at different sections of the source and try to give an account of its morphological evolution. In Section 2 we present radio observations and data reduction procedure in detail. In Section 3 we present our results which are followed by a discussion in Section 4. Final conclusions are drawn in Section 5.

All absolute quantities in this paper were calculated for a $\Lambda$CDM universe with $H_0$ = 70 km s${^{-1}}$Mpc${^{-1}}$ and $\Omega_{\rm{m}}$ = 0.3 and $\Omega_{\rm{\Lambda}}$ = 0.7.  The host galaxy is referred to as CGCG 292-057 and the related radio source is called J1159+5820 throughout the paper. Using the host galaxy redshift, the conversion scale translates to 1.045 kpc/\arcsec,  used in the paper.

\begin{figure*}
  \centering
  \includegraphics[scale=1.03]{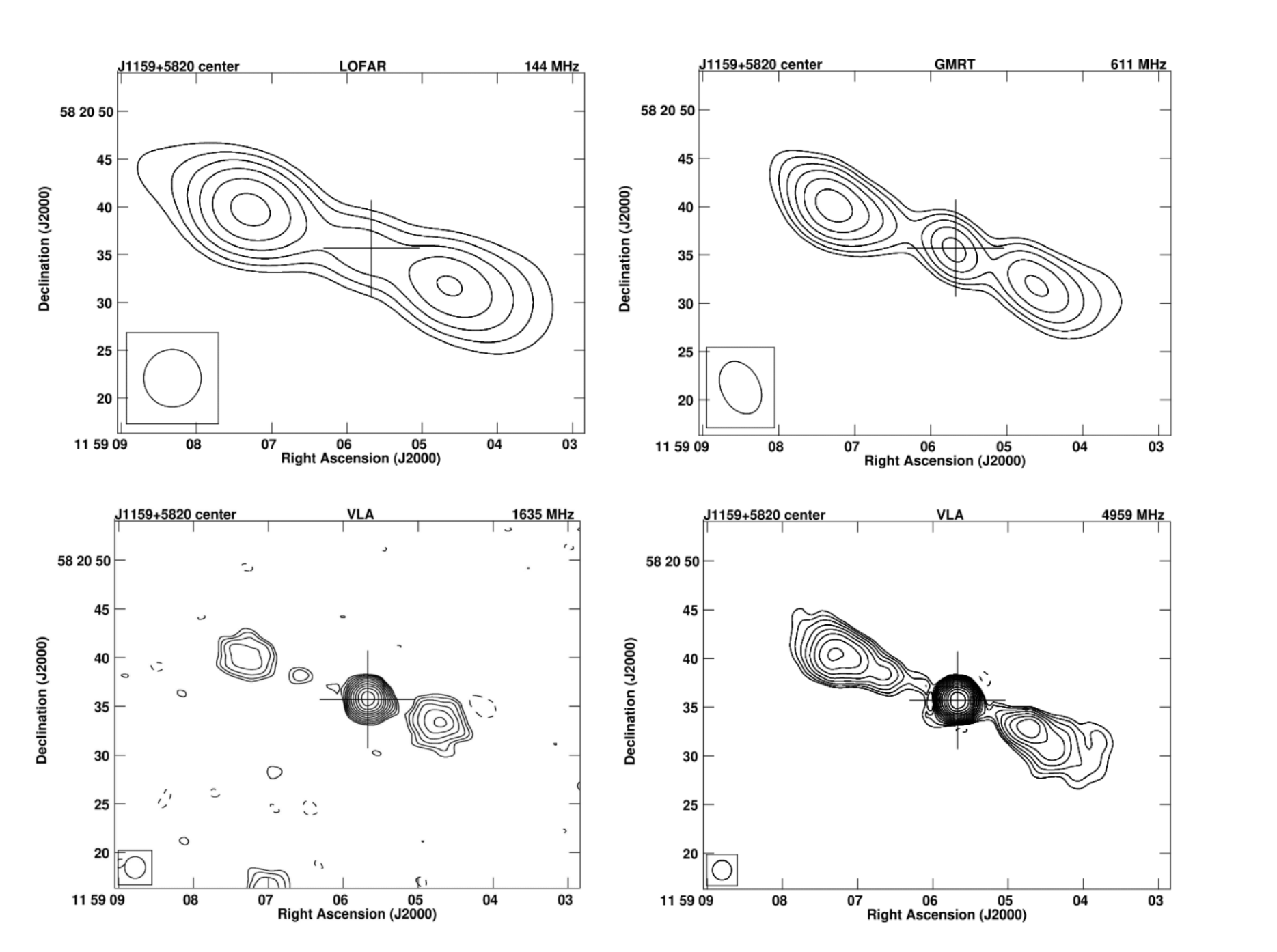}
  \caption{Contour maps of the  central part of J1159+5820 (within the host galaxy), obtained from LOFAR, GMRT, and the VLA observations at 144, 611, 1635, and 4959 MHz (see also right panel of Fig.~\ref{fig1}). The contour levels are spaced by a factor of $\sqrt{2}$ and the first contour is at 3 $\times$ rms level (rms values are given in Table~\ref{table1}). The first negative 3 $\times$ rms level is marked with a dashed contour. The relative sizes of the beam are indicated by the ellipse at the bottom left corner of each image. The position of the core is marked with a cross.}   
  \label{fig2}
\end{figure*}

\section{Radio Observations}

 To understand the radio morphology of J1159+5820, high-resolution and multifrequency radio maps of the diffuse lobes, center, and the overall structure of the radio source were obtained from dedicated observations with the Giant Metrewave Radio Telescope (GMRT) at low-frequency, the Karl G. Jansky Very Large Array (VLA) at high-frequency, and publicly available survey data. The shortest baselines from both telescopes could successfully sample the largest scales of the source. In the following subsections, we describe in detail the surveys used, the observation schemes, and the data reduction techniques used to obtain GMRT and the VLA maps.

\begin{table*}
\caption{Details of GMRT and the VLA dedicated observations of J1159+5820 analyzed in this paper.}
\label{tab:landscape}
\setlength{\tabcolsep}{5.8 pt}
\renewcommand{\arraystretch}{0.73}
\begin{tabular}{cccccccccc}
\hline
Freq. &Telescope & Proposal code   & Date of observation & TOS & Bandwidth & Channel   &Size & PA & rms \\
(MHz) & &  &  & (hrs) & (MHz)&  &(\arcsec$\times$\arcsec) & (\degree)   & (mJy/beam)  \\
(1) & (2) & (3) & (4) & (5) & (6)& (7) & (8) & (9) & (10)  \\
\hline
147 & GMRT  &  20$\_$009  & 2011 May 14  & 7.30 & 16.6 & 256 & 29.1$\times$16.5  & 28.0 & 1.279 \\
\\
 240 & GMRT     & 18$\_$026  & 2010 May 21  & 5.26 & 16.0 & 128 &13.0$\times$10.3  & 27.4   &0.843 \\
 \\
323  & GMRT      &  20$\_$009   & 2011 Aug 22  &   5.55 & 33.3 & 256   &15.0$\times$7.8  & 20.7 & 0.229 \\
\\
611 & GMRT     & 18$\_$026  & 2010 May 21  & 5.26 &  33.3  & 512   &5.6$\times$4.0  & 23.4 & 0.067   \\
\\
1570 & VLA L band C-conf       & 10B-127	  & 2010 Nov 02  & 0.53 & 256 & 128   & 18.4$\times$14.0 & 63.1 & 0.204  \\
\\
 1635  & VLA L band A-conf     & 10C-130  &2011 Jul 08   &  0.47 & 256 &  128  & 2.2$\times$1.0 & -52.9 & 0.040  \\
 \\
4959  &VLA C band CNB-conf      &   10C-130 & 2011 Feb 20  &  2.25   & 256  &  128&2.0$\times$1.4 & -85.2  & 0.011  \\
\\
4959   &VLA C band C-conf$^*$     & 10B-127   & 2010 Nov 02  & 1.40  & 256 &  n/a  &    n/a           & n/a  &  n/a \\
\\
4959 & VLA C band D-conf$^*$    & 11B-051     & 2011 Nov 08  & 1.35  & 256 &  n/a   &    n/a &  n/a  &  n/a  \\       

\hline
\end{tabular}
\begin{tablenotes}\footnotesize
\item[*] \textit{*}  The VLA C band observations in C and D configurations were imaged simultaneously with a resulting resolution of 15\farcs6$\times$11\farcs9, PA of -4\fdg9 and the rms of 0.06 mJy/beam.
\end{tablenotes}
\label{table1}
\end{table*}

\begin{table*}
\caption{Flux densities of different components of J1159+5820.}
\label{tab:landscape}
\begin{tabular}{cccccccccc}
\hline
 Freq. & core &  inner eastern lobe & inner western lobe &  eastern primary lobe   & western primary lobe &  total structure &  References\\
 (MHz) & (mJy) & (mJy) & (mJy)  & (mJy)  & (mJy)  & (mJy)  &    \\
(1) & (2) & (3) & (4) & (5) & (6)& (7) & (8)  \\
\hline
54 & & & & & & 3530 $\pm$ 700 & I \\
 144  & 2.6 $\pm$ 0.2 & 45.8 $\pm$ 2.3 & 40.3 $\pm$ 2.0 & 404 $\pm$ 65 & 556 $\pm$ 94   &1660 $\pm$ 95 & II \\

147 & & & &  & & 1677 $\pm$ 294     & This paper \\

240 & & & & 258 $\pm$ 48 & 338 $\pm$ 73 & 1264 $\pm$ 223$^b$  & This paper   \\
323 & 5.2 $\pm$ 0.7 & 28.8 $\pm$ 2.9$^a$  & 42.6  $\pm$ 4.5$^a$ & 276 $\pm$  49 & 375 $\pm$  71 & 1220 $\pm$ 197$^b$ & This paper\\
611  & 4.8 $\pm$ 0.4 & 15.6 $\pm$ 1.3$^a$ & 13.2  $\pm$ 1.1$^a$ & 168 $\pm$ 30 & 225 $\pm$ 45  & 798 $\pm$ 98$^b$ & This paper\\
1400 & 1.8 $\pm$ 0.2 & 3.9 $\pm$ 0.5 &  1.8  $\pm$ 0.4  & &      & 338 $\pm$ 57$^c$ & III, IV \\

1570 & & &  & 84 $\pm$ 14  & 139  $\pm$ 25  &  & This paper\\
1635 & 4.1 $\pm$ 0.2  & 1.3 $\pm$ 0.4  &  1.6  $\pm$ 0.5 &  &   &  & This paper\\

3000 & 5.6 $\pm$ 0.5 & 7.1 $\pm$ 1.5 & 6.3 $\pm$ 1.2  & &    & & V\\
4959 & 7.3 $\pm$ 0.4 & 3.2 $\pm$ 0.8 & 2.4 $\pm$ 0.6 & 51 $\pm$ 9$^d$ & 65  $\pm$  13$^d$ & 166 $\pm$ 17$^d$ &   This paper\\
8440 &9.2 $\pm$ 0.6 & & & &     & & VI \\
\hline
\end{tabular}
\begin{tablenotes}\footnotesize
\item[*] \textit{References} - (I) \cite{2021A&A...648A.104D}; (II) \cite{2022A&A...659A...1S}; (III) \cite{1995ApJ...450..559B}; (IV) \cite{1998AJ....115.1693C}; (V) \cite{2020PASP..132c5001L}; (VI) \cite{2007ApJS..171...61H}; (a) flux of inner lobes with shorter baselines flagged to get the compact emission; (b) flux of total structure with longer baselines flagged to obtain extended emission, (c) total flux taken from NVSS; and (d) values from C band C and D configuration.
\end{tablenotes}
\label{table2}
\end{table*}

\subsection{GMRT observations}
\label{sec:gmrt observations} 

J1159+5820 was observed with the GMRT in four frequency bands centered at 147, 240, 323, and 611 MHz. All observations were performed within projects 18$\_$026 and 20\_009 in May 2010 and between May and August 2011, respectively (details are given in Table~\ref{table1}). The usual scheme of observations of looping the phase calibrator with the target source was adopted with the flux density calibrators observed for $\sim$10-20 mins at the beginning and at the end of observations. For the observation of the target source centered at 147 MHz and 323 MHz, 3C\,286 was observed as both the phase and flux density calibrator. At 147 MHz, 3C\,286 as the phase calibrator was observed for 5 mins alternating with the target source in the loop. At 323 MHz, 3C\,286 was observed as the flux density calibrator and also as the phase calibrator for 4 mins between cycles. Observations at 240 MHz and 611 MHz were  conducted in dual-frequency mode (\citealt{1991CSci...60...95S}).
This mode allows the recording of data at two frequency bands  simultaneously but with a single polarisation for each band only, thereby reducing the sensitivity by a factor of $\sim$1.41. In this case, flux density calibrators 3C\,48 and 3C\,147 were observed at the beginning and 3C\,286 at the end of the observations. The phase calibrator for this set of observations was J1313+675, which was observed for 5 mins in a loop with the target source. 

The data reduction for all observed frequencies was completed using the Source Peeling and Atmospheric Modeling (SPAM) pipeline (\citealt{2009A&A...501.1185I}; \citealt{2014ascl.soft08006I}). It is a Python-based extension to AIPS (\citealt{2003ASSL..285..109G}), and is aimed at reducing high-resolution, low-frequency radio interferometric observations. It contains direction-dependent calibration, modeling, and corrections for dispersive phase delay, mainly of ionospheric
origin.  The data were corrected for strong radio frequency interference (RFI) and then standard flux density, phase, and bandpass calibrations were applied to the source. Cleaning was performed using the Cotton-Schwab algorithm (\citealt{1984AJ.....89.1076S}; \citealt{1999ASPC..180..357C}), which is a variant of CLEAN deconvolution
(\citealt{1974A&AS...15..417H}; \citealt{1980A&A....89..377C}) that allows for the simultaneous deconvolution of multiple facets, using a different dirty beam for each
facet. The resultant images were then primary beam corrected, using the AIPS task PBCOR.

\begin{figure*}
  \centering
  \includegraphics[scale=0.90]{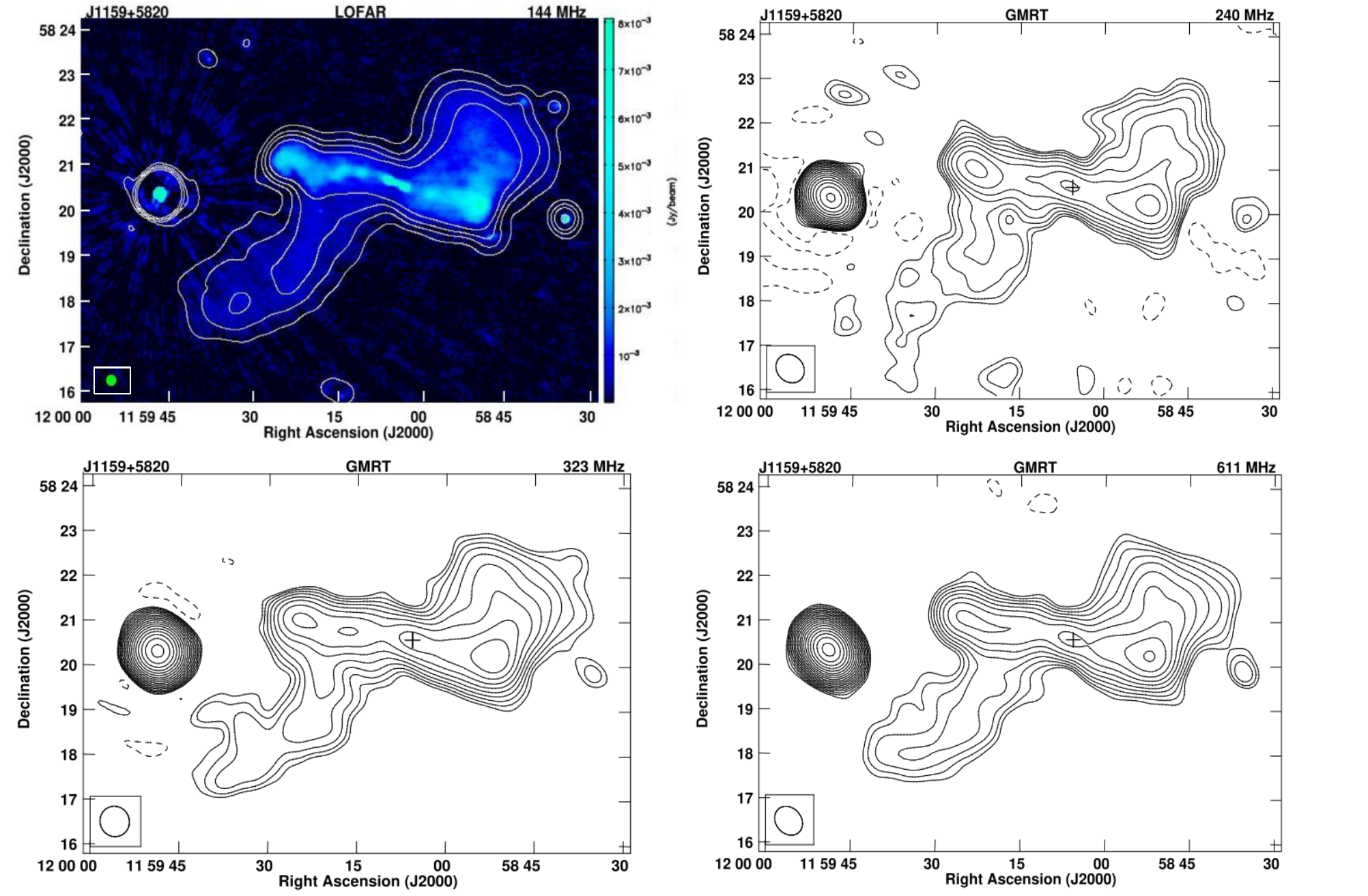}
  \caption{Low-frequency maps of J1159+5820 obtained from LOFAR and GMRT observations at 144, 240, 323, and 611 MHz. The contour levels are spaced by a factor of $\sqrt{2}$ and the first contour is at 3 $\times$ rms level (rms values given in Table~\ref{table1}). The first negative 3 $\times$ rms level is marked as a dashed contour. The relative sizes of the beam are indicated by the ellipse at the bottom left corner of each image. The position of the core is marked with a cross. The top left corner map is LOFAR 144 MHz 6\arcsec\ high-resolution map (given in blue) overlayed with LOFAR 20\arcsec\ low-resolution contours. The bright source in the background is an independent compact steep-spectrum (CSS) source.}  
  \label{fig3}
\end{figure*}

\subsection{VLA observations}

J1159+5820 was observed with the VLA between November 2010 and November 2011 in C and L bands using various array configurations. C and D configurations allowed for proper sampling of the largest scale emission from the source.
The more compact structure of the inner lobes could be efficiently studied with the use of CNB configuration at C band and A configuration at L band, both 
allowing for detecting structures as extended as 30\arcsec at resolutions of few arcseconds.
The details of the observations are presented in Table~\ref{table1}. 
All observations were performed in single scheduling blocks consisting of calibrator (both flux and phase) and source scans. 

The data were reduced with the Common Astronomy Software Applications (CASA) package\footnote{http://casa.nrao.edu} (\citealt{2022PASP..134k4501C}).
All data affected by shadowed antennas or RFI were flagged manually.
The radio source 3C\,286 was used for the flux density scale in all observations, and phase calibration was done with the use of 
the radio source J1148+5924 in the case of L-band observations in A-configuration and the radio source J1219+4829 for all other observations.

All data sets, as presented in Table~\ref{table1}, were calibrated separately, resulting in five calibrated data sets (for each band and configuration). To deconvolve the point spread function (PSF) of the interferometer from the dirty map, we used the CASA task {\sc tclean}, which performs the Clark CLEAN algorithm. The two C-band observations in compact configurations C and D were merged prior to cleaning to increase the uv-coverage of the data and allow for a more precise deconvolution. To produce the best quality images we used the robust weighting of the data (robust parameter set to one) and the multi-scale CLEAN. Due to a very strong point source just east of J1159+5820 (details on this source are given in section 3.5), in all of the obtained maps, severe interference from its sidelobes was still present. Therefore, the maps were self-calibrated both in phases (initial runs) and in amplitudes (final runs). The final images were restored with elliptical beams  and then convolved into a circular beam. This allowed for better identification of components of the radio source, 
as well as to improve the signal-to-noise ratio of the faint extended emission.

\subsection{Data extracted from surveys}

Low-frequency surveys used in the multifrequency analysis of the source include LOw-Frequency ARray (LOFAR; \citealt{2013A&A...556A...2V}) LBA Sky Survey (LoLSS; \citealt{2021A&A...648A.104D}) and LOFAR  Two-metre Sky Survey (LoTSS; \citealt{2022A&A...659A...1S}). LoLSS maps the entire northern sky in the frequency range 42-66 MHz, with a resolution of 15\arcsec\,at an average rms of 1 mJy/beam. LoTSS Data Release 2 is an ongoing  sky survey ranging from 120-168 MHz with a 6\arcsec\,resolution and median rms sensitivity of 83 $\micro$Jy/beam and a 20\arcsec\,resolution with a median rms sensitivity of 95 $\micro$Jy/beam.
The VLA surveys that were used for high-frequency analysis include Faint Images of the Radio Sky at Twenty centimeters (FIRST; \citealt{1995ApJ...450..559B}), NRAO
VLA Sky Survey (NVSS; \citealt{1998AJ....115.1693C}), and the VLA Sky Survey
(VLASS; \citealt{2020PASP..132c5001L}). FIRST at 1.4 GHz has an rms of 0.15 mJy/beam with a resolution of 5\arcsec\, and covers regions of sky in the north and south galactic cap. The NVSS is a 1.4 GHz continuum survey covering the entire sky north of -40$\degree$ declination with a resolution of 45\arcsec. The rms brightness fluctuations are about 0.45 mJy/beam in total power (Stokes I). VLASS is an all-sky survey above declination -40$\degree$. It operates within a range of 2-4 GHz with an angular resolution of ~2.5\arcsec\,and sensitivity of 69 $\micro$Jy/beam. In Table~\ref{table2}, we show the flux density values obtained from dedicated observations and surveys.   

\section{Results}

\subsection{Radio morphology}

In the large-scale emission seen in the low-frequency maps presented in Fig.~\ref{fig1} and Fig.~\ref{fig3}, J1159+5820 displays a pair of well-defined axially symmetric lobes labeled as eastern primary lobe (EPL) and western primary lobe (WPL), accompanied by two extended and diffuse wings, labeled as north western wing (NWW) and south eastern wing (SEW; marked in Fig.~\ref{fig1}). The LOFAR 20\arcsec $\times$ 20\arcsec\,resolution map overlaid on the LOFAR 6\arcsec $\times$ 6\arcsec\,contour map at 144 MHz, given in Fig.~\ref{fig3} (upper left panel), shows the extent of the diffuse SEW elongated much further in the southeastern direction with the high-resolution map clearly showing jet bending in the EPL, unseen before. The WPL displays a hotspot-like feature around  RA = $11\fhr58\fmin50\fs6$, Dec = $+$58\degr20\arcmin04\arcsec (J2000.0), with a spectral index value of about 0.5, which is typical for a hotspot. In the GMRT 610 MHz and high-frequency VLA maps in Fig.~\ref{fig2}, the central structure is clearly resolved. The core, along with the inner lobes, prominently show an FRII-type mini-double. Thus, the target source exhibits a double-double morphology. J1159+5820 is also close to a bright compact steep spectrum (CSS) source, which is discussed in detail in Section 3.5.

J1159+5820 resembles an FRII-type winged RG, however, the radio power of the primary lobes $P_{\rm1400MHz} \simeq 2 \times 10^{24}{\rm W\, Hz^{-1}}$, places the source at the FRI/FRII transition boundary. The primary lobes are symmetric with respect to the radio core and the source displays an almost X-symmetric structure with the two low surface brightness wings emerging outward from the high surface brightness lobes with the estimated inclination of the structure of $\sim$80$\degree$. This structure shares a strong similarity to the source NGC 507 (\citealt{2011A&A...526A.148M}; \citealt{2022A&A...661A..92B}). WISEA J130225+514911 is another source found in the LoTSS DR2 survey that also demonstrates  morphological resemblance to our target source but at a much larger distance of $z_{\rm{phot}}$ = 0.9.  The host galaxy is seen face-on and its centre coincides with the radio core. This makes the linear size of the source from the eastern to the western primary lobe of $\sim$330 kpc and of the inner pair of lobes $\sim$30 kpc. The formation of distinct outer and inner lobes traced at least two different epochs of the jet cycle, with the outer lobes being formed at a different cycle of jet activity than the inner pair,  hinting at  recent changes in the level of nuclear activity of the source.

\begin{figure}
 
  \includegraphics[scale=0.57]{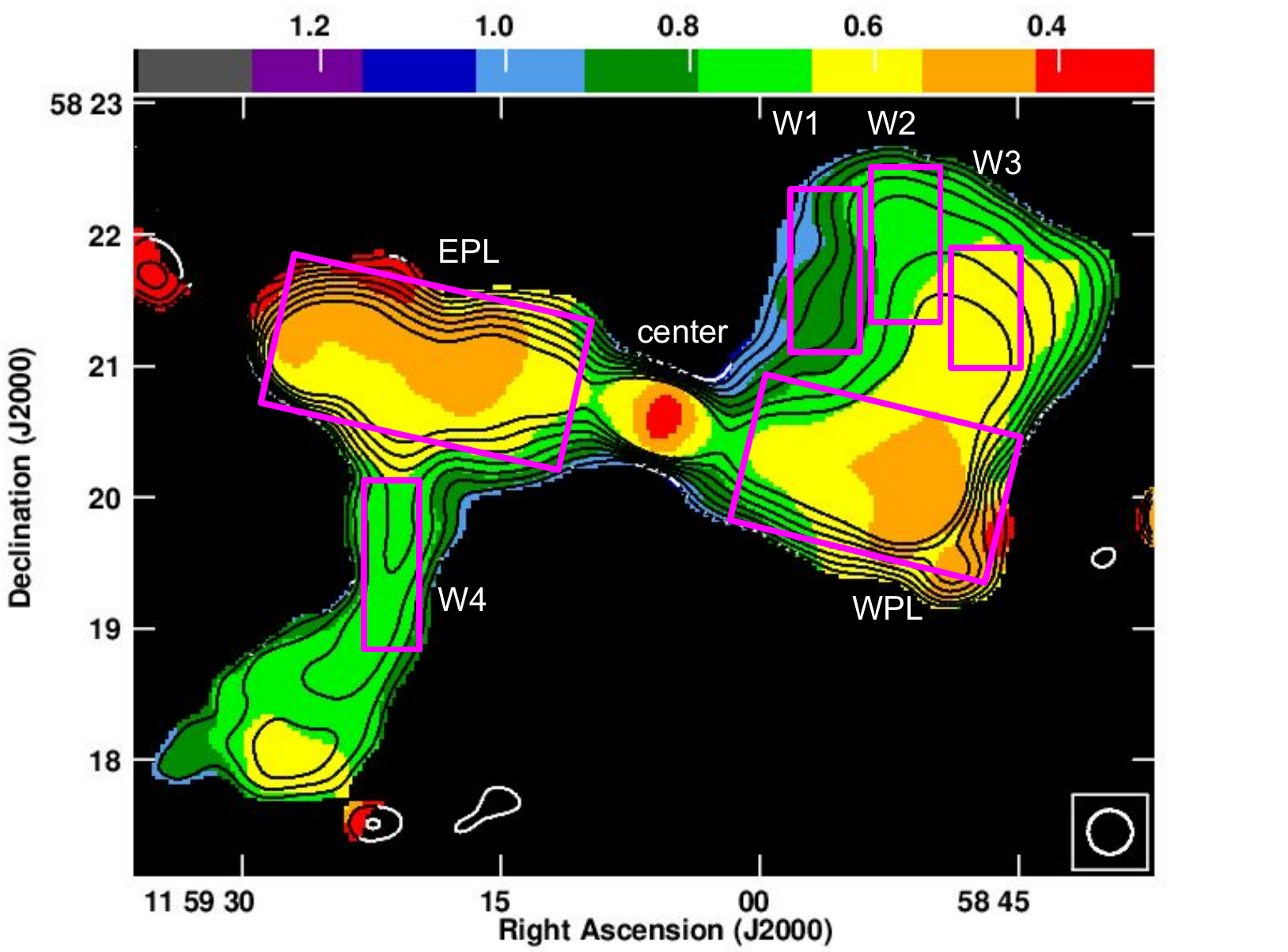}
  \caption{ The spectral index map between 144 MHz and 4959 MHz. The VLA 4959 MHz contours are spaced by a factor of $\sqrt{2}$ with the first contour starting at 3 $\times$ rms level. The size of the beam is indicated by a circle in the bottom right corner of the image. Four different regions
in the wings (W1-W4), as well as the regions of the primary lobes (WPL and EPL) are marked with magenta rectangles. The center region includes both the core and the inner lobes. In further sections, these regions are used to perform multifrequency ageing analysis.}
  \label{fig4}
\end{figure}

\subsection{Spectral index analysis}

 The profile of radio spectra conveys important information regarding the evolution of plasma in the source. A steep spectrum hints at old plasma that is devoid of fresh energetic electrons, unlike a flatter spectrum where fresh particles might still be injected. In radio lobes, synchrotron and inverse-Compton losses cause steepening of the spectra over time, with more energetic electrons losing energy faster than the less energetic ones. In this case, the spectral index (SI) provides crucial information about the past activity of a radio source. The SI map of J1159+5820 was obtained using the LOFAR 144 MHz and the VLA 4959 MHz map. The VLA map was first tapered to a resolution similar to the LOFAR 144 MHz map and was then convolved to 20\arcsec x 20\arcsec\,, that is, the resolution of the original LOFAR map. AIPS task {\sc hgeom} and {\sc comb} were used respectively to align the geometry of both of the maps and to finally produce the spectral index map as shown in Fig.~\ref{fig4}. All the other pre-described GMRT and VLA maps were similarly transformed before using them for further spectral analyses of the different regions of J1159+5820. The SI map ($S_{\nu} \propto \nu^{-\alpha}$) reveals a relatively flat core and steep spectra wings. The spectrum in the lobes is quite different from that of the wings, with the north-western wing showing a strong gradient going from steeper to flatter spectra along a clockwise direction. This shows the concentration of energetic electrons in the WPL, in comparison to the wings. In the SEW, the spectra are steep, but with much less gradient, indicating the presence of old and low energetic electrons. The EPL shows relatively flat spectra, similar to WPL, although it flattens further out than EPL. There is a small region of flat spectra around the edges of the eastern and western primary lobe, which might be a result of an image artifact. The steeper spectra next to the inner lobes are caused by the backward-flowing lobe plasma, where aged plasma accumulates.  The SEW looks narrower than the counter NWW, as the former is much weaker and has a relatively lower surface brightness. The strong gradient of the flux density contours in the EPL, as seen in Fig.~\ref{fig4}, could also be a hint of missing flux in the region between the eastern primary lobe and the bright background source. Nevertheless, the lack of significant steepening of the spectral index throughout the edges of the source confirms that  no large-scale emission was lost in the high-frequency VLA observations.

To acutely understand the movement of plasma across the wings, they were further divided into four separate regions based on the same/similar SI values (W1-W4; Fig.~\ref{fig4}). As the gradient shows a trend in relative plasma ages, it will help to estimate when particles were last accelerated in these regions. It is also evident that the total flux of the primary lobes is mostly affected by the brightest compact nodes of the flat-spectrum emission, and as such is not representative  of the true lobe age. To estimate the ages of the primary lobes it would be much proper to measure the spectral age with the particle injection model fitted to the inner edge of the primary  lobes (IEPL) and to the EPL and WPL. The spectral age of the EPL/WPL will give the mean age of the lobes, whereas the spectral age of the IEPL will constitute the oldest plasma from primary lobes and should give the true age of the lobes. However, since the IEPL might include contamination
from the old plasma from the wings, it could return an age comparatively older than the true age of the primary lobes. Therefore, ages estimated from IEPL and EPL/WPL will be more likely the upper and lower constraints for the ages of the primary lobes. Further analyses are presented in more detail in  the next sections.

\begin{figure}
\includegraphics[width=0.38\textheight]{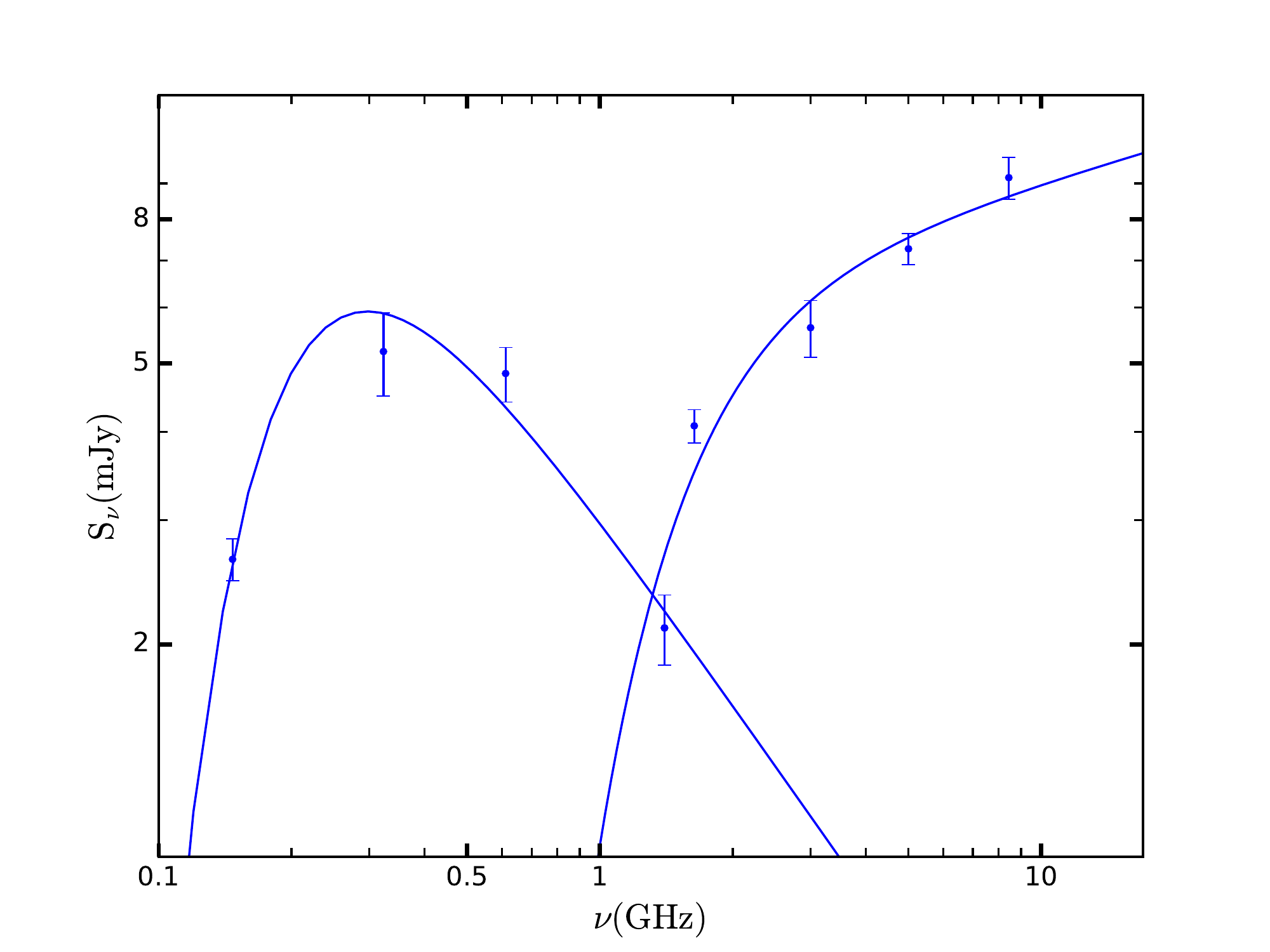}
\caption {Radio spectra of the core with the double homogeneous FFA model fit as discussed in section 3.3.}
\label{fig5}
\end{figure}

\begin{figure}
\includegraphics[width=0.38\textheight]{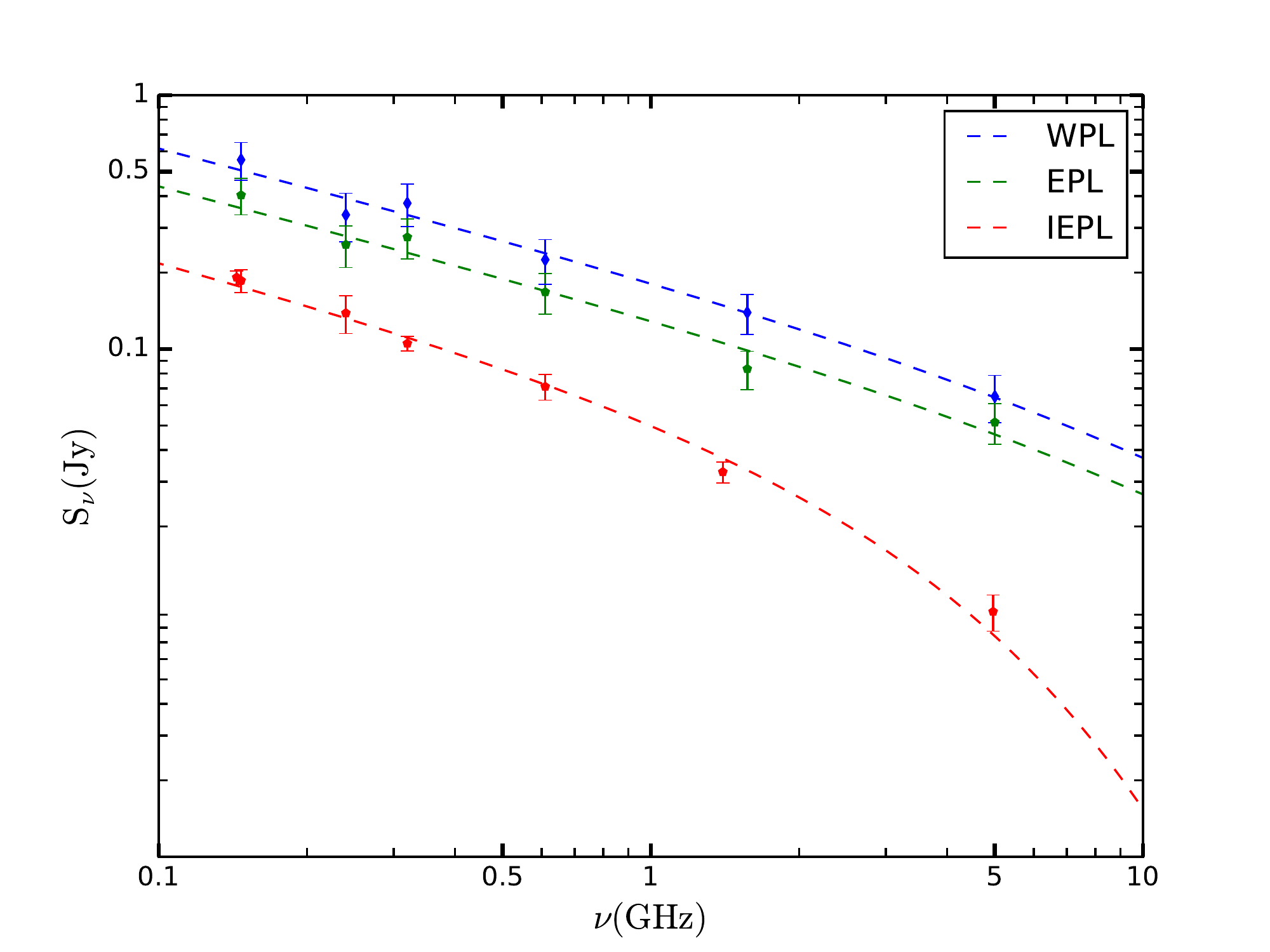}
\caption{Radio spectra of the eastern and western primary lobes (EPL; WPL) fitted with the CI model and the inner edge of the primary lobes (IEPL) fitted with the JP model.}
\label{fig6}
\end {figure}

\begin{figure}
\includegraphics[width=0.38\textheight]{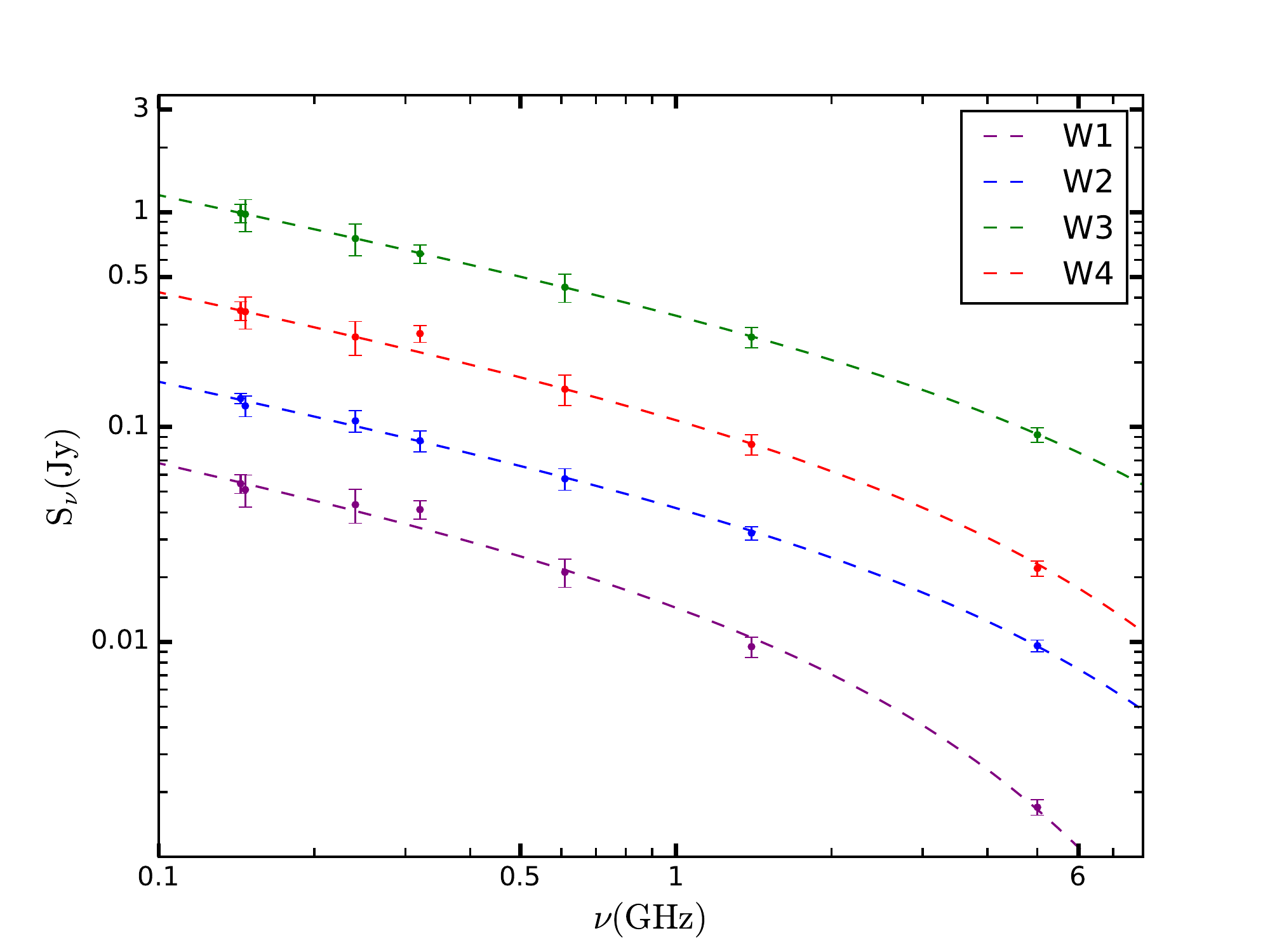}
\caption{Radio spectra of regions W1, W2, W3, and W4 (marked in Fig. 3) fitted with the JP model. The spectra of particular regions are arbitrarily shifted in the ordinate axis to give the appropriate picture of the curvature of the spectra.}
\label{fig7}
\end {figure}

The flux densities of all components of the source are listed (Table~\ref{table2} and Table~\ref{table4}) and have been used to model the individual components comprising the pair of inner lobes, the core, the primary lobes, and different sections of the wings. The resulting spectra fitted using different particle injection models to the primary lobes and wings, and absorption models to the core are displayed in Figs.~\ref{fig5}-\ref{fig7}.

\subsection{Radio core}

The core spectrum is plotted from 144 MHz to 8440 MHz in Fig.~\ref{fig5}, using the flux density values given in Table~\ref{table2}. The spectrum steeply rises from 144 MHz and then falls rapidly by 1400 MHz; it then rises steadily again around 1635 MHz and continues to rise till 8440 MHz. The first low-frequency turnover, below 500 MHz, implies a low-frequency absorption. Such low-frequency absorption is also seen in the case of CSS sources. The turnover frequency observed is closely associated with the absorption of synchrotron radiation in the source, which is either in the form of synchrotron self-absorption (SSA) or free-free absorption (FFA) (\citealt{1966AuJPh..19..195K}). To understand this mechanism we  tested the SSA and FFA models to describe the core spectra similar to \cite{2015ApJ...809..168C}.  All models gave a similar fit including the homogeneous FFA model which provided the least $\chi^2_{\rm{red}}$ value. This particular model describes an external homogeneous ionized screen around a synchrotron emitting plasma that attenuates the radiation. The absorbing medium can be described by 

\begin{equation}
\\\\\\\ S_{\nu} = a{\nu}^{-\alpha}e^{-\tau_{\nu}}
\end{equation}

where a and $\alpha$ are the amplitude and spectral index of the synchrotron spectrum and $\tau_{\nu}$ is the optical depth. The optical depth is given by $\tau_{\nu}=(\frac{\nu}{\nu_{\rm{t}}})^{-2.1}$, where $\nu_{\rm{t}}$ is the turnover frequency at which the optical depth is equal to 1. 
A single model fit to the spectrum was not possible, hence two separate homogeneous FFA models were fit, the first one starting from 144 MHz to 1400 MHz and the second fit starting from 1400 MHz to 8440 MHz. This led to two independent model fits for the core spectra as given in Fig.~\ref{fig5}. The fitting parameters are given in Table~\ref{table3}. The values of the turnover frequency and spectral index in the second model fit are similar to the ones obtained by \cite{2001ApJ...552..120L} in their study of the core spectrum of the radio source B0313-192, which is also associated with a disc galaxy. \\

\begin{table}
\caption{Model fit parameters for the core with double homogeneous FFA model}

\label{tab:landscape}
\setlength{\tabcolsep}{3.3 pt}
\begin{tabular}{cccccccccc}
\hline
Freq. range (GHz) &  a & $\nu_{\rm{t}}$ (GHz) & $\alpha$ & $\chi^2_{\rm{red}}$  \\
(1)& (2) & (3) & (4) & (5)\\
 \hline
 0.14-1.4 &3.06 $\pm$ 0.22  & 0.198 $\pm$ 0.01  & 0.89 $\pm$ 0.13 & 0.87  \\
1.4-8.4& 5.98 $\pm$ 1.47 & 1.305 $\pm$ 0.17   & -0.18 $\pm$ 0.14  &  4.91  \\

\hline
\end{tabular}
\label{table3}
\end{table}

\begin{table}
\caption{Flux densities of wing regions and the inner edge of primary lobes.}
\label{tab:landscape}
\renewcommand{\arraystretch}{1.2}
\setlength{\tabcolsep}{2.8pt}

\begin{tabular}{cccccccccc}
\hline
Freq. & W1 & W2  & W3 & W4  & IEPL \\
(MHz) & (mJy)  & (mJy)  & (mJy) & (mJy) & (mJy)\\
(1)& (2) & (3) & (4) & (5) & (6)\\
 \hline
 144 & 54.5 $\pm$ 4.3  & 135.9 $\pm$ 7.6  & 107.4 $\pm$ 6.3 & 79.2 $\pm$  7.9  &  191.5 $\pm$ 11.5  \\
147  &  51.1  $\pm$ 7.6 & 125.4 $\pm$ 13.7  & 89.1 $\pm$ 10.5 & 57.4 $\pm$  9.8  &  186.2 $\pm$ 19.0 \\
240  & 43.5  $\pm$ 7.2 & 106.8 $\pm$ 12.1  & 74.8 $\pm$  9.4 & 68.6 $\pm$ 12.3  & 138.6 $\pm$ 23.4 \\
323   &  41.3  $\pm$ 7.2 &  86.3 $\pm$ 9.7  & 64.5 $\pm$ 7.8 & 68.7 $\pm$ 7.5   & 105.2 $\pm$ 7.1  \\
611  &    21.1 $\pm$ 3.1 &  57.4 $\pm$ 6.2  & 46.7 $\pm$ 5.2  & 33.2 $\pm$ 5.3 &  71.1 $\pm$ 8.2  \\
1400    &  9.5 $\pm$ 1.7 & 32.1 $\pm$ 2.3  & 23.1 $\pm$ 2.0 &  17.4 $\pm$ 2.0 &  32.8 $\pm$ 3.1  \\
4959   & 1.7  $\pm$ 0.3 & 9.6 $\pm$ 0.6  &  9.5 $\pm$ 0.6 &  5.2 $\pm$ 0.4 &  9.2 $\pm$ 1.5 \\

\hline
\end{tabular}
\label{table4}
\end{table}

\begin{table}
\begingroup
\caption{Break frequency values for the primary lobes and wing regions}

\label{tab:landscape}
\setlength{\tabcolsep}{13.5pt}
\renewcommand{\arraystretch}{1.5}
\begin{tabular}{cccccccc}
\hline
Lobe/Region & Model & Break Freq. & $\chi^2_{\rm{red}}$  \\
 &  & ($\nu_{\rm{br}}$ in GHz) & \\
(1) & (2) & (3) & (4) \\
\hline
 EPL& CI    & $8.4  ^{+>100}_{-5.9} $    & 0.65 \\
 WPL& CI    & $8.1  ^{+>100}_{-6.9}$     &  0.30 \\
IEPL & JP &  $4.8  ^{+>100}_{-0.7}$ & 1.16\\
 W1&  JP    & $3.8  ^{+1.9}_{-0.8} $ & 0.45 \\
 W2&  JP  & $8.5  ^{+6.5}_{-1.7} $  & 0.14 \\
 W3&  JP    & $13.2  ^{+>100}_{-2.2} $ & 1.11  \\
 W4&   JP & $7.7 ^{+6.5}_{-2.7}$  & 2.31 \\
 
\hline
\end{tabular}
\label{table5}
\endgroup
\end{table}

According to \cite{1998PASP..110..493O}, the turnover frequency and the projected linear size of the radio source can give crucial information regarding the source's linear evolution. This correlation is given by

\begin{equation}
\\\\\\\\\ log(\nu_{\rm{t}}) = -0.21(\pm0.05)-0.65(\pm0.05)log(l)
\end{equation}

where $l$ is the linear size of the source in  kpc and $\nu_{\rm{t}}$ is the turnover frequency in GHz. Making use of two different turnover frequencies from the model fit, we estimate two different linear sizes, $\sim$5 kpc corresponding to the first turnover at 0.2 GHz and $\sim$320 pc corresponding to the second turnover at 1.3 GHz. The electron density in these regions was calculated using the emission measure for free-free absorption (\citealt{1998PASP..110..493O}) as follows:
 
\begin{equation}
\\\\\\\ (n_{\rm{e}})^2l \simeq 3.05 \times 10^6  \tau  \left( \frac{\rm{T}}{10^4\rm{K}} \right)^{1.35}
\left( \frac{\nu}{1 \rm{GHz}} \right)^{2.1} {\rm{cm^{-6}}} {\rm{pc}}
\end{equation}

 where $n_{\rm{e}}$ is the electron density, $\tau$ is the optical depth at frequency $\nu$ in GHz, and T is the temperature in K. Using the two different turnover frequency values, we arrive at two different electron density values, that is for the turnover at 0.2 GHz,  $n_{\rm{e}}$ is $\sim$5 $\rm{cm^{-3}}$ and for the turnover corresponding to 1.3 GHz,  $n_{\rm{e}}$ is $\sim$126 $\rm{cm^{-3}}$. The two different electron density values most likely correspond to different regions of the central nuclear region through which the radiation is absorbed as it passes. It initially passes through the denser inner parsec-scale region of $\sim$320 pc with a higher electron density of about 126 $\rm{cm^{-3}}$, which is comparable to the value of $\sim$200 $\rm{cm^{-3}}$ that \cite{2005MNRAS.362..931E} obtained from the central kpc region of 3C\,293,  before entering a less opaque region of the ISM at around 5 kpc, where  $n_{\rm{e}}$ drops to about 5 $\rm{cm^{-3}}$.

 \subsection{Lobes and Wings}

\subsubsection{Source Energetics}

 The spectrum of J1159+5820 was fitted with the Jaffe \& Perola \citep[JP;][]{1973A&A....26..423J}  and continuous injection \citep[CI;][]{1970ranp.book.....P} models for calculating the radiative losses that compute time evolution using the initial power-law energy distribution. This is performed using the emission spectrum of particles specified by their injection spectral index ($\alpha_{\rm{inj}}$) distributed isotropically in pitch angle relative to the direction of the magnetic field. In the case of both the eastern and western primary lobe, the CI model provided the best fit to the data with minimum  $\chi^2_{\rm{red}}$ values. The NWW was divided into three different regions based on the spectral gradient using the SI map (see Fig.~\ref{fig4}) and the SEW was made into a single vertical region due to a lack of clear gradient across the lobe. All the regions of the northern wing and the southern wing were best fit with the JP model, possibly due to the lack of any compact hotspots along with the presence of steep spectra in the wings. The same is the case of the IEPL region. The SYNAGE package (\citealt{1996PhDT........92M}) was used to fit the CI and JP models to the radio spectra. The best fit models are given in Fig.~\ref{fig6} and Fig.~\ref{fig7} and the fit parameters are given in Table~\ref{table5}. With the application of the JP and CI model, the following assumptions were made: (i) the particles follow a constant power-law energy distribution without any re-acceleration of radiating particles after entering the lobes, (ii) the magnetic field lines are intertwined and the field strength is constant throughout the energy-loss process, and (iii) the time of isotropization is short compared to the radiative lifetime for the pitch angles for the injected particles. We address that the assumptions made here are an idealization of the physical processes that are actually taking place in the lobes, for instance, the supposition of a constant magnetic field. However, since the dominant loss mechanism in these regions is inverse-Compton, therefore it is reasonable to assume a constant magnetic field here.

\subsubsection{Primary Lobes and Wings}
  The break frequency values obtained for the EPL and WPL are given in Table~\ref{table5}, for a fixed value of $\alpha_{\rm{inj}}$= 0.5. Since from the pre-fitting results $\alpha_{\rm{inj}}$ was close to 0.5, the $\alpha_{\rm{inj}}$ therefore was fixed at that particular value. The $\nu_{\rm{br}}$  values for both primary lobes obtained via the CI model give comparable results with $\nu_{\rm{br}}$ lying slightly above the highest radio frequency data presented in the paper. As visible in Fig.~\ref{fig4}, the close vicinity of the central region, IEPL, is characterized by a steeper spectrum than the rest of the primary lobes, and its emission is best fitted with the JP model (see Fig.~\ref{fig6}). The JP model fit for the wing regions W1, W2, W3, and W4 is given in Table~\ref{table5}. The spectra of the primary lobes, as given in Fig.~\ref{fig6}, appear flatter than the steep spectra wings, given in Fig.~\ref{fig7}, implying the absence of recent particle injection processes in the wings. In the regions W1, W2, and W3, we see a clear gradient in the break frequency values. which seems consistent with the spectral index map. This shows that the oldest plasma is in region W1 followed by regions W2 and W3. This directs the movement of the flow of plasma and the jet in the clockwise direction in the NWW. 

\begin{figure}
\includegraphics[width=0.39\textheight]{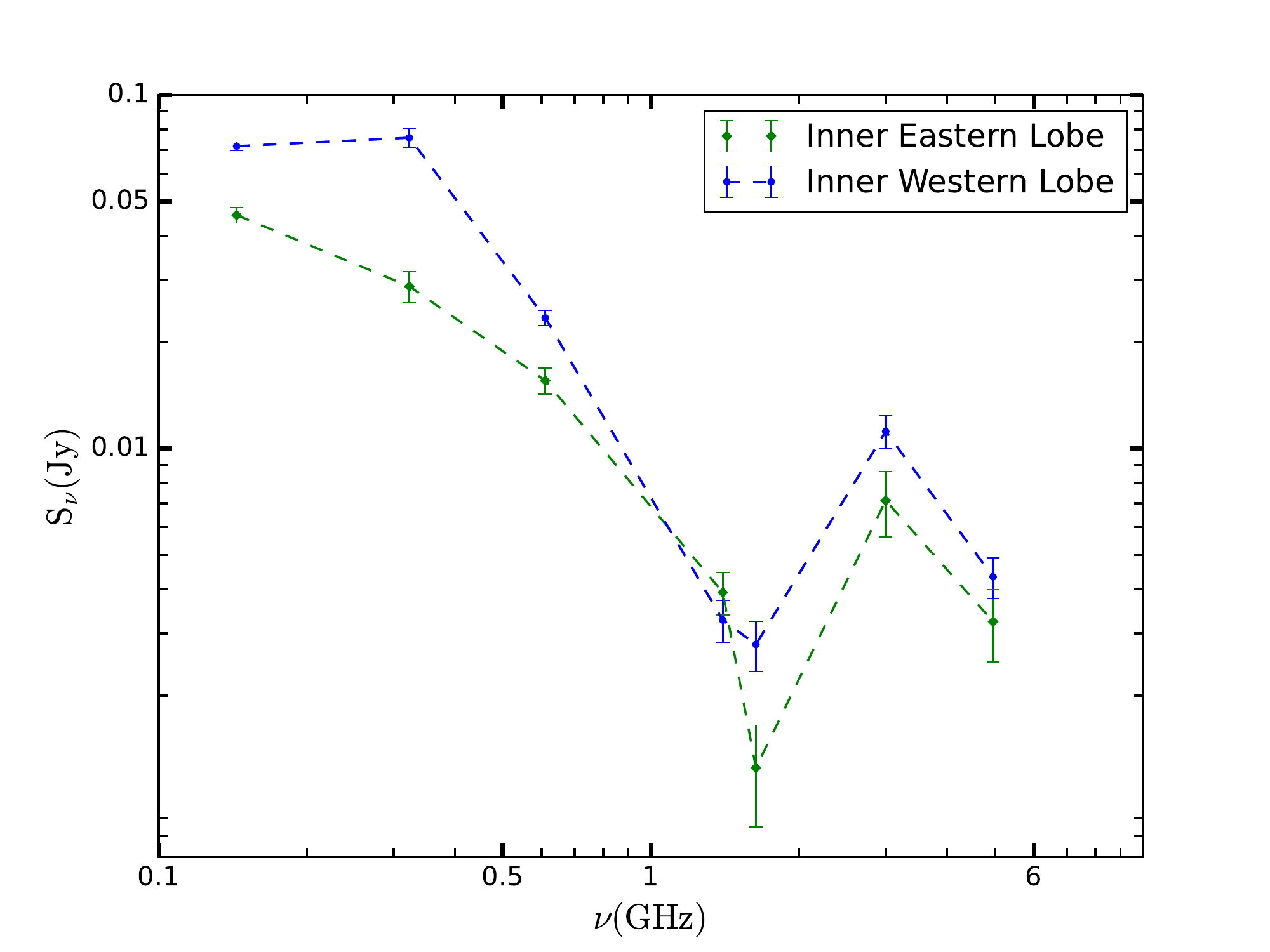}
\caption {Radio spectra of the inner lobes. The spectrum of the inner western lobe is shifted arbitrarily in the ordinate axis to give the appropriate picture of the spectra.}
\label{fig8}
\end{figure}

We estimate the magnetic field strength and synchrotron age for the primary lobes, IEPL, and the regions W1, W2, W3, and W4  of the wings. The spectral age is calculated using the following formula: 

\begin{equation}
 \tau_{\rm{rad}}=50.3 \frac{B_{\rm{eq}}^{0.5}}{B_{\rm{eq}}^2 + B_{\rm{CMB}}^2} (\nu_{\rm{br}}(1+z))^{-0.5} {\mathrm{Myr}}
\end{equation}

$B_{\rm{CMB}} = 0.318(1 + z)^2$ is the magnetic field strength analogous to the cosmic microwave background radiation at the redshift of the target. Here B, the magnetic field strength of the lobes, and $B_{\rm{CMB}}$ are expressed in units of nT, $\nu_{\rm{br}}$ the spectral break frequency in GHz above which the radio spectrum steepens from the initial power-law spectrum is given by $\alpha_{\rm{inj}}$=($\gamma-1$)/2.
The magnetic field was calculated using the minimum energy arguments following Longair (2011). The magnetic field strength can be directly estimated when a radio source is detected at radio and X-ray frequencies simultaneously. X-ray emission in the lobes of RGs most likely originate from inverse-Compton scattering between the same relativistic electrons that produce the observed radio synchrotron radiation, and the CMB (\citealt{1979MNRAS.188...25H}). For the calculation of magnetic field, with the lack of adequate X-ray observations, using the equipartition assumption is the most general practice. However, it is also often found that in the lobes of active FRII sources, the magnetic field is generally within a factor of a few (2--3) lower than those implied by equipartition (e.g. \citealt{2002ApJ...581..948H}; \citealt{2005ApJ...626..733C}; \citealt{2005ApJ...622..797K}; \citealt{2007ApJ...668..203M}; \citealt{2017MNRAS.467.1586I}; \citealt{2018MNRAS.474.3361T}). For inactive (outer) lobes of double-double radio galaxies that were analyzed by \cite{2019MNRAS.486.3975K}, it was found that  their magnetic field values were close to the ones obtained by equipartition. In addition, \cite{1999A&A...344....7P} showed (see their fig. 4) how the radiative lifetime depends on the ratio of magnetic field $B/B_{\rm{eq}}$. They proved that for $0.5 < B_{\rm{eq}}/B_{\rm{CMB}} < 2$, deviations from equipartition have a small effect on the computed lifetime if $B\leq2·B_{\rm{eq}}$. Furthermore, in the case of our target, inverse-Compton losses to the CMB seem to dominate over synchrotron losses. Therefore, the equipartition assumption to calculate the magnetic field is reasonable for the analysis that we conducted. Our assumptions included the cutoff frequencies of $\nu_{\rm{min}}$ = 10 MHz and $\nu_{\rm{max}}$ = 100 GHz, the filling factor of 1, and the pure electron-positron plasma. The volume of the lobes and different regions of the wings was estimated using a cylindrical shape. This led to the magnetic field strength values for the eastern and western primary lobe as B = 0.10 ± 0.01 nT and B = 0.11 ± 0.01 nT respectively. The spectral ages of the eastern primary lobe and western primary lobe are calculated as 39.4 Myrs and 41.7 Myrs respectively. For IEPL the B = 0.14 ± 0.01 nT and the spectral age is 58.2 Myrs.

For the NWW regions W1, W2, and W3, the B = 0.11 ± 0.01, 0.17 ± 0.01, 0.16 ± 0.01 nT respectively, for the SEW region W4 the B = 0.14 ± 0.01 nT. Taking into account the break frequencies
calculated by the model and the magnetic field strength estimated
above from a cylindrical geometry, the spectral ages of NWW
are 61.8 Myrs, 45.1 Myrs and 36.0 Myrs for regions W1, W2, and W3, respectively, for SEW region W4 is 45.6 Myrs. This shows that the oldest plasma is in the region W1, followed by W2 and W3. The ages of W2 and W4 are similar, which is in accordance with the SI map in Fig.~\ref{fig3}. The decreasing gradient in the SI map and the spectral ages of different wing regions in the NWW confirm the clockwise movement of plasma. The mean spectral age of the primary lobes is $\sim$40 Myrs, however, the actual age of these lobes considering the oldest plasma from IEPL is $\sim$60 Myr (as detailed in Section 3.2).

\begin{figure}
 
  \includegraphics[scale=0.46]{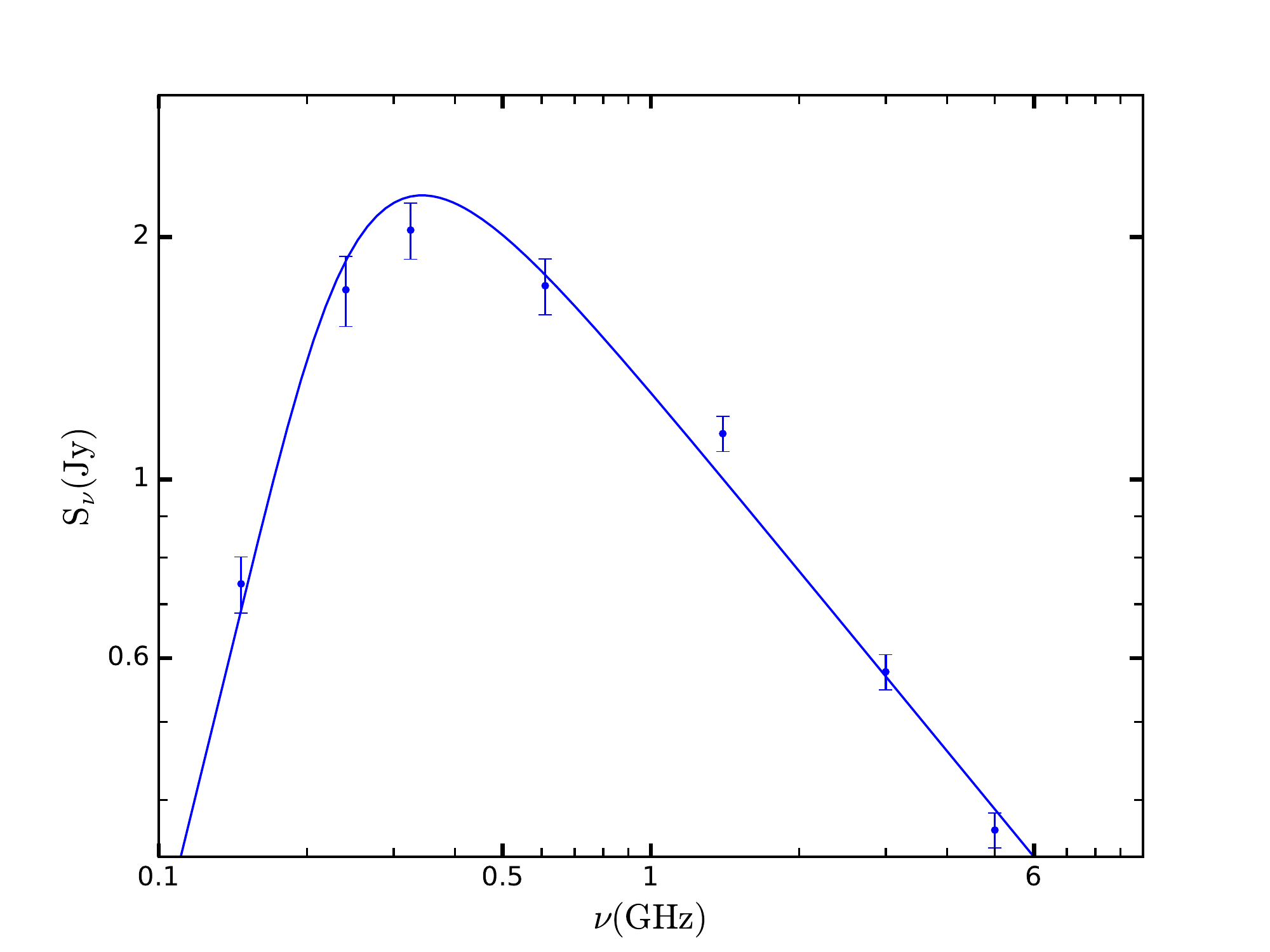}
  \caption{Radio spectrum of the CSS source next to J1159+5820 modeled using synchrotron self-absorption.  }   
  \label{fig9}
\end{figure}

\subsubsection{Inner Lobes}

The flux density of the inner lobes is plotted in Fig.~\ref{fig8}. The spectrum shows a convex pattern with an unusual bump at 3000 MHz and a slight resemblance to some GPS sources described in \cite{2017ApJ...836..174C}. Upon fitting, the CI and JP models for the data from 144 MHz to 4959 MHz, to the inner lobes, the break frequency values obtained from the fit were unreasonable, giving spectral ages much older than the steep spectra wings. This discrepancy could arise due to strong absorption in the ISM as the inner lobes expand through them. Therefore, we could not provide accurate spectral ages of the inner lobes. However, for dynamical age, considering a lobe expansion speed of 0.2c (\citealt{2003PASA...20...19M}), we estimate the inner lobes should not be younger than $\sim$0.2 Myrs.

\subsection{Background CSS source}

The CSS source located at RA = $11\fhr59\fmin48\fs7$, Dec = $+$58\degr20\arcmin20\arcsec (J2000.0), next to the eastern primary lobe of the target source is at a redshift of z = 1.27 (\citealt{2016MNRAS.459..820T}). The measured flux density values at different frequencies are given in Table~\ref{table6}. Its radio spectrum fitted with the SSA model is given in Fig.~\ref{fig9}. Also, to check for any discrepancies with the flux density values of the maps, we fitted the CSS source next to the target with the SSA model which returned a reasonable fit with a turnover frequency of $\sim$270 MHz as given in Fig.~\ref{fig9}.

\begin{table}
\begingroup
\caption{Flux density values for the CSS source}

\label{tab:landscape}
\setlength{\tabcolsep}{3 pt}
\renewcommand{\arraystretch}{1}
\begin{tabular}{cccccccc}
\hline
 Freq. (MHz) & 147 & 240 & 323 & 611 &1400 &3000 & 4959\\
 \hline
 Flux density (Jy)  & 0.74 & 1.72 & 2.04 & 1.74 & 1.14 & 0.58 & 0.37 \\

 Error (Jy)  &$\pm$0.06 & $\pm$0.17 & $\pm$0.16 & $\pm$0.14 & $\pm$0.06 & $\pm$0.03 & $\pm$0.02\\
 
\hline
\end{tabular}
\label{table6}
\endgroup
\end{table}

\section{Discussion}

To give an account of the extended radio morphology of J1159+5820, we proceed by discussing some possible scenarios:
\\

\subsection{Hydrodynamic Backflow}
The radio jets from luminous RGs usually evolve to become inversion symmetric about their host galaxies. In some cases however, the backward-flowing plasma from the primary lobes interacts with the ISM of the host galaxy and starts expanding laterally. When there is a misaligned hot gas halo around the host galaxy, it can bend the backward-flowing plasma in opposite directions creating secondary wings (\citealt{1984MNRAS.210..929L}; \citealt{2020MNRAS.495.1271C}). These wings are therefore a result of a single AGN outburst, with the backward-flowing plasma from the jets being diverted along the direction of the steepest pressure gradient of the surrounding medium. However, if this is the case for our target source, it will be difficult to explain wings that are larger than the primary lobes. This is  due to the fact that energetic jets are usually expected to drive through the surrounding medium supersonically, and wings will have to expand completely under buoyancy, which is usually  subsonic. The spectral index map also shows a gradation in the distribution of plasma in the wings (especially in the NWW), instead of a perfectly steep spectra expected from a backward-flowing cocoon, due to quasi-simultaneous evolution. The south eastern wing is very extended in the low-frequency maps; however, in the high-frequency VLA map (see Fig.~\ref{fig3}), no such emission is found implying a much steeper emission. In the Chandra X-ray studies conducted by \cite{2020ApJ...905..148B}, the presence of an X-ray halo around the host was not observed, but this could result from a limited sensitivity of Chandra to large-scale diffuse emission.\\

\subsection{Spin flip of SMBH}

Galaxy mergers most likely take place in galaxy groups and clusters. CGCG\,292-057 does not reside in such an environment, but upon closer inspection  $\sim$7 galaxies were found within a radius of around 1 Mpc with similar redshift.  Although the number of galaxies within a megaparsec distance to our host is few, this nevertheless does not rule out the possibility that there might be other galaxies at a similar redshift not discovered or cataloged by surveys. However, at an angular separation of 17\farcm5 from CGCG 292-057, an edge-on galaxy MCG +10-17-123 at RA = $12\fhr00\fmin04\fs0$, Dec = $+$58\degr36\arcmin20\arcsec (J2000.0) with z = 0.0541,  obscured by an S-shaped dust lane is located. This disturbed galaxy has a comparable projected angular size, brightness, and redshift to our target galaxy. To explore further the possibility of whether these two galaxies might have interacted via a close flyby event in the past, we considered a typical velocity of 250 km/s (\citealt{1983MNRAS.205..605B}) for each galaxy and a distance of 500 kpc (assuming the flyby might have occurred midway to the current distance separating them). The time to reach the current separation was calculated as $\sim$2 Gyrs.\\
  The binary SMBHs formed through galaxy mergers mostly coalesce following a regime of slow approach as a result of dynamical friction and gravitational radiation. As discussed by \cite{2002Sci...297.1310M}, galaxy mergers can lead to a dramatic change in the SMBH spin axis direction, subsequently leading to a sudden flip in the direction of the associated jet, post-merger of the system. This mechanism will lead to a distinct X-shaped morphology, e.g. J1513+2607 (\citealt{Springmann2006HOSTGO}). The two pairs of jets here would be formed from two different episodes of nuclear activity, with the new lobes formed before the older lobes disappear. To produce such a pronounced X-shaped morphology would require a large spin-flip angle, which would lead to the creation of active primary lobes post-spin flip of SMBH. However, in the case of J1159+5820, we do not see two prominent pairs of lobes passing symmetrically through the core of the associated galaxy; instead, they appear to be emerging rather radially from the primary lobes.

\subsection{Independent jets from Dual AGN}
Furthermore,  as first proposed by \cite{2005MNRAS.356..232L}, there might exist a class of binary co-orbiting SMBHs with independent jets undergoing a merger, e.g. 3C 75 (\citealt{2005LRR.....8....8M}), which due to poor resolution or orientation effects appear to generate an X-shaped morphology. From the optical maps of CGCG 292-057, we cannot resolve a second nuclear core, and neither in the radio maps and/or X-ray maps there is a distinct dual radio core. This either hints that the binary SMBH pair have merged or has fallen below kpc scale separation. A third possibility could be that the secondary SMBH is not active at the present time of observation. All of these scenarios are difficult to be explored currently in the confines of this paper.\\

\subsection{Jet Precession}
As discussed in \cite{1980Natur.287..307B}, a natural way for a black hole to undergo precession at a significant rate is while orbiting in a binary system. As mentioned by \cite{1992A&A...254...96K}, the apparent shape of any jet generated by the black holes will be modified by the binary's orbital motion, which can motivate the helical motion of jets or precession. These wiggles in jets are observed in the quasar 1928+738 and are successfully modeled by \cite{1993ApJ...409..130R}. In 3C\,273, \cite{2000A&A...360...57R} showed that the precession of the inner jets results from tidal perturbation of the accretion disk produced by the non-coplanar orbit of the secondary black hole. The exact morphology according to \cite{2002Sci...297.1310M} however would depend on the timescale of jet axis reorientation. A slow realignment would lead to an FRI source with S-shaped morphology, with rapid realignment producing an intermediate-luminosity X-shaped source, with radio power near the FRI/FRII break. Relatively slowly precessing jets can also lead to an X-shaped morphology in projection, as shown by hydrodynamical simulations performed by \cite{2020MNRAS.499.5765H}.

In the optical study conducted by \cite{2012MNRAS.422.1546K}, the SDSS nuclear spectrum revealed a doubly peaked profile, which hints at two possibilities: it could be either due to Narrow Line Region (NLR) kinematics or as a result of an SMBH binary. The NLR can be both associated with AGN outflows and galaxy mergers (\citealt{2018ApJ...867...66C}). The lack of an optically detectable companion galaxy here can imply that we are observing the post-merger stage where the black hole separation has shrunk below the kpc scale. There still remains the possibility of the jet precession being induced by a tilted accretion disk, as suggested by \cite{1990A&A...229..424L}. As elaborated by \cite{1997MNRAS.292..136P}, the accretion disk can become wrapped due to nonuniform irradiation from AGN, causing instabilities of the accretion disk and leading to precession. But a wrapped accretion disk cannot explain the NLR kinematics. But since we cannot resolve two separate optical cores, it is hard to rule out one of the above precession models. In the LOFAR map given in Fig.~\ref{fig3} (upper left panel), one can clearly see the wiggles or waves in the eastern jet, which can hint at an underlying precession in progress. \\

\subsection{Jet-shell Interaction}
In the jet-shell interaction model for the formation of X-shaped radio galaxies proposed by \cite{2012RAA....12..127G}, the radio morphology arises from the interaction of jets with gaseous shells in a merged galaxy system. According to the model, the merger between galaxies leads to the formation of a sequence of shells, and the wings are formed as a result of the interruption faced by the jets trying to pass through rotating shells. These temporary deflections can be triggered by a lighter jet colliding with a few times wider massive cloud. This model can well account for the observed Z-shaped wings, since also in the optical map (given in Fig.~\ref{fig1}) there seems to be a shell-like structure associated with the host galaxy almost exactly along the direction of the northwestern wing. There might be a possibility that the shells are formed as the galaxy moves during the merger from an inclined position almost clockwise in the sky plane to its current face-on position, and thereby the jets coming from SMBH interact with the shells and undergo bending until the galaxy settles to its current orientation.   \\

Among all the models discussed here, it might be possible that more than one phenomenon is at play behind the origin of the radio morphology. However, the model that most likely explains the relative spectral ages in the different regions of the wings and primary lobes is the precessing jet model. It is favored by the decreasing gradient of the spectral ages in the different regions of the NWW, directing at a clockwise movement of plasma driven by a fast reorientation of the jets, and it is also further supported by the wiggles in the jet, distinctly visible in the EPL of the LOFAR 144 MHz map (see Fig.~\ref{fig3}).

\section{Conclusion}

CGCG 292-057 is an exceptional source that demonstrates multiple stages of galaxy evolution at once. The evidence of a past merger, peculiar X-shaped morphology, and an AGN rebirth with a new pair of radio lobes makes it a great source for studying the evolution of galaxies.

To analyze the source properties, we conducted multifrequency observations using GMRT and the VLA. The conclusion we draw is that the most likely explanation for the X-shaped morphology is the jet reorientation following the orbital motion of a binary SMBH pair causing a rapid jet precession in the timescale of a few million years.
\begin{itemize}

\item We have supplemented the flux density measurements from our observational data with a large number of radio sky surveys that helped us to build the spectra from 54 MHz to 8440 MHz. Using the above, we modeled the lobes and different sections of the wings and were able to deduce the break frequency and conduct an ageing analysis.

\item The core was modeled using a double homogeneous FFA model as a single FFA could not fit the spectra. This is most likely caused due to the presence of two different ionizing screens around the synchrotron emitting relativistic plasma, which could be due to the result of the past merger activity. 

\item The inner pair of lobes produced unrealistic break frequency values with various particle injection models, ultimately giving much larger ages than the steep spectra wings. We reason that the strong absorption caused during the expansion of the inner pair of lobes inside the ISM of the host galaxy could cause such an unusual behavior of the radio flux. 

\item The spectral age of primary lobes was found to be between 40 Myrs and 58 Myrs. We also found the ages of the different sections of the steep spectra wings to be 61.8, 45.1 and 36.0 Myrs for regions W1, W2, and W3, respectively, and 45.6 Myrs for region W4. This decreasing gradient of the spectral ages of the wing regions agrees well with that of the SI map.

\end{itemize}
The multiwavelength study of this one-of-kind post-merger double-double source with a peculiar morphology opens up a new window for thorough investigations of the origins of such class of radio galaxies. Future observations with ALMA will allow the detection of cool molecular gas and possible outflows in the galaxy, which can shed light on the unusual behavior of the core and the inner lobes. This can subsequently lead to a better understanding of the effects of AGN feedback on the host galaxy and SMBH-galaxy coevolution.

\section{Acknowledgements}
The authors thank the anonymous Reviewer for her/his valuable comments and suggestions. We thank Subhrata Dey and Dorota  Kozieł-Wierzbowska for their help. MJ acknowledges access to the SYNAGE software kindly provided by M. Murgia. We thank the GMRT staff for their assistance with observations. The GMRT is a national facility operated by the NCRA, TIFR. The National Radio Astronomy Observatory is a facility of the National Science Foundation operated under a cooperative agreement by Associated Universities Inc. This research work was partially supported by the National Science Center OPUS-15 grant, nr UMO-2018/29/B/ST9/01793. AM also acknowledges the Jagiellonian University grant: N17/MNS/000055.\\

LOFAR data products were provided by the LOFAR Surveys Key Science project (LSKSP; https://lofar-surveys.org/) and were derived from observations with the International LOFAR Telescope (ILT). LOFAR (van Haarlem et al. 2013) is the LowFrequency Array designed and constructed by ASTRON. It has observing, data processing, and data storage facilities in several countries, which are owned by various parties (each with their own funding sources), and which are collectively operated by the ILT foundation under a joint scientific policy. The ILT resources have benefited from the following recent major funding sources: CNRS-INSU, Observatoire de Paris and Université d'Orléans, France; BMBF, MIWF-NRW, MPG, Germany; Science Foundation Ireland (SFI), Department of Business, Enterprise and Innovation (DBEI), Ireland; NWO, The Netherlands; The Science and Technology Facilities Council, UK; Ministry of Science and Higher Education, Poland; The Istituto Nazionale di Astrofisica (INAF), Italy.

This research made use of the Dutch national e-infrastructure with support of the SURF Cooperative (e-infra 180169) and the LOFAR e-infra group. The Jülich LOFAR Long Term Archive and the German LOFAR network are both coordinated and operated by the Jülich Supercomputing Centre (JSC), and computing resources on the supercomputer JUWELS at JSC were provided by the Gauss Centre for Supercomputing e.V. (grant CHTB00) through the John von Neumann Institute for Computing (NIC).

This research made use of the University of Hertfordshire high-performance computing facility and the LOFAR-UK computing facility located at the University of Hertfordshire and supported by STFC [ST/P000096/1], and of the Italian LOFAR IT computing infrastructure supported and operated by INAF, and by the Physics Department of Turin university (under an agreement with Consorzio Interuniversitario per la Fisica Spaziale) at the C3S Supercomputing Centre, Italy. 

The National Radio Astronomy Observatory running the VLA is a facility of the National Science Foundation operated under a cooperative agreement by Associated Universities, Inc.

This research has made use of BASS DR3 images. BASS is a key project of the Telescope Access Program (TAP), which has been funded by the National Astronomical Observatories of China, the Chinese Academy of Sciences (the Strategic Priority Research Program “The Emergence of Cosmological Structures” Grant No. XDB09000000), and the Special Fund for Astronomy from the Ministry of Finance. The BASS is also supported by the External Cooperation Program of Chinese Academy of Sciences (Grant No. 114A11KYSB20160057), and Chinese National Natural Science Foundation (Grant No. 12120101003, No. 11433005).

The Pan-STARRS1 Surveys (PS1) and the PS1 public science archive have been made possible through contributions by the Institute for Astronomy, the University of Hawaii, the Pan-STARRS Project Office, the Max-Planck Society and its participating institutes, the Max Planck Institute for Astronomy, Heidelberg and the Max Planck Institute for Extraterrestrial Physics, Garching, The Johns Hopkins University, Durham University, the University of Edinburgh, the Queen's University Belfast, the Harvard-Smithsonian Center for Astrophysics, the Las Cumbres Observatory Global Telescope Network Incorporated, the National Central University of Taiwan, the Space Telescope Science Institute, the National Aeronautics and Space Administration under Grant No. NNX08AR22G issued through the Planetary Science Division of the NASA Science Mission Directorate, the National Science Foundation Grant No. AST-1238877, the University of Maryland, Eotvos Lorand University (ELTE), the Los Alamos National Laboratory, and the Gordon and Betty Moore Foundation.

\section*{Data Availability}

The data underlying this paper will be shared on reasonable request to the corresponding author.



\bibliographystyle{mnras}
\bibliography{cgcg} 

\begin{thebibliography}{}
\makeatletter
\relax
\def\mn@urlcharsother{\let\do\@makeother \do\$\do\&\do\#\do\^\do\_\do\%\do\~}
\def\mn@doi{\begingroup\mn@urlcharsother \@ifnextchar [ {\mn@doi@}
  {\mn@doi@[]}}
\def\mn@doi@[#1]#2{\def\@tempa{#1}\ifx\@tempa\@empty \href
  {http://dx.doi.org/#2} {doi:#2}\else \href {http://dx.doi.org/#2} {#1}\fi
  \endgroup}
\def\mn@eprint#1#2{\mn@eprint@#1:#2::\@nil}
\def\mn@eprint@arXiv#1{\href {http://arxiv.org/abs/#1} {{\tt arXiv:#1}}}
\def\mn@eprint@dblp#1{\href {http://dblp.uni-trier.de/rec/bibtex/#1.xml}
  {dblp:#1}}
\def\mn@eprint@#1:#2:#3:#4\@nil{\def\@tempa {#1}\def\@tempb {#2}\def\@tempc
  {#3}\ifx \@tempc \@empty \let \@tempc \@tempb \let \@tempb \@tempa \fi \ifx
  \@tempb \@empty \def\@tempb {arXiv}\fi \@ifundefined
  {mn@eprint@\@tempb}{\@tempb:\@tempc}{\expandafter \expandafter \csname
  mn@eprint@\@tempb\endcsname \expandafter{\@tempc}}}

\bibitem[\protect\citeauthoryear{{Adelman-McCarthy} et~al.,}{{Adelman-McCarthy}
  et~al.}{2008}]{2008ApJS..175..297A}
{Adelman-McCarthy} J.~K.,  et~al., 2008, \mn@doi [\apjs] {10.1086/524984},
  \href {https://ui.adsabs.harvard.edu/abs/2008ApJS..175..297A} {175, 297}

\bibitem[\protect\citeauthoryear{{Balasubramaniam}, {Stawarz}, {Marchenko},
  {Sobolewska}, {Cheung}, {Siemiginowska}, {Thimmappa}  \&
  {Kosmaczewski}}{{Balasubramaniam} et~al.}{2020}]{2020ApJ...905..148B}
{Balasubramaniam} K.,  {Stawarz} L.,  {Marchenko} V.,  {Sobolewska} M.,
  {Cheung} C.~C.,  {Siemiginowska} A.,  {Thimmappa} R.,   {Kosmaczewski} E.,
  2020, \mn@doi [\apj] {10.3847/1538-4357/abc4e2}, \href
  {https://ui.adsabs.harvard.edu/abs/2020ApJ...905..148B} {905, 148}

\bibitem[\protect\citeauthoryear{{Bean}, {Ellis}, {Shanks}, {Efstathiou}  \&
  {Peterson}}{{Bean} et~al.}{1983}]{1983MNRAS.205..605B}
{Bean} A.~J.,  {Ellis} R.~S.,  {Shanks} T.,  {Efstathiou} G.,   {Peterson}
  B.~A.,  1983, \mn@doi [\mnras] {10.1093/mnras/205.3.605}, \href
  {https://ui.adsabs.harvard.edu/abs/1983MNRAS.205..605B} {205, 605}

\bibitem[\protect\citeauthoryear{{Becker}, {White}  \& {Helfand}}{{Becker}
  et~al.}{1995}]{1995ApJ...450..559B}
{Becker} R.~H.,  {White} R.~L.,   {Helfand} D.~J.,  1995, \mn@doi [\apj]
  {10.1086/176166}, \href
  {https://ui.adsabs.harvard.edu/abs/1995ApJ...450..559B} {450, 559}

\bibitem[\protect\citeauthoryear{{Begelman}, {Blandford}  \& {Rees}}{{Begelman}
  et~al.}{1980}]{1980Natur.287..307B}
{Begelman} M.~C.,  {Blandford} R.~D.,   {Rees} M.~J.,  1980, \mn@doi [\nat]
  {10.1038/287307a0}, \href
  {https://ui.adsabs.harvard.edu/abs/1980Natur.287..307B} {287, 307}

\bibitem[\protect\citeauthoryear{{Bessiere}, {Tadhunter}, {Ramos Almeida}  \&
  {Villar Mart{\'\i}n}}{{Bessiere} et~al.}{2012}]{2012MNRAS.426..276B}
{Bessiere} P.~S.,  {Tadhunter} C.~N.,  {Ramos Almeida} C.,   {Villar
  Mart{\'\i}n} M.,  2012, \mn@doi [\mnras] {10.1111/j.1365-2966.2012.21701.x},
  \href {https://ui.adsabs.harvard.edu/abs/2012MNRAS.426..276B} {426, 276}

\bibitem[\protect\citeauthoryear{{Brienza} et~al.,}{{Brienza}
  et~al.}{2022}]{2022A&A...661A..92B}
{Brienza} M.,  et~al., 2022, \mn@doi [\aap] {10.1051/0004-6361/202142579},
  \href {https://ui.adsabs.harvard.edu/abs/2022A&A...661A..92B} {661, A92}

\bibitem[\protect\citeauthoryear{{Browne} et~al.,}{{Browne}
  et~al.}{2003}]{2003MNRAS.341...13B}
{Browne} I.~W.~A.,  et~al., 2003, \mn@doi [\mnras]
  {10.1046/j.1365-8711.2003.06257.x}, \href
  {https://ui.adsabs.harvard.edu/abs/2003MNRAS.341...13B} {341, 13}

\bibitem[\protect\citeauthoryear{{Callingham} et~al.,}{{Callingham}
  et~al.}{2015}]{2015ApJ...809..168C}
{Callingham} J.~R.,  et~al., 2015, \mn@doi [\apj]
  {10.1088/0004-637X/809/2/168}, \href
  {https://ui.adsabs.harvard.edu/abs/2015ApJ...809..168C} {809, 168}

\bibitem[\protect\citeauthoryear{{Callingham} et~al.,}{{Callingham}
  et~al.}{2017}]{2017ApJ...836..174C}
{Callingham} J.~R.,  et~al., 2017, \mn@doi [\apj]
  {10.3847/1538-4357/836/2/174}, \href
  {https://ui.adsabs.harvard.edu/abs/2017ApJ...836..174C} {836, 174}

\bibitem[\protect\citeauthoryear{{Capetti}, {Zamfir}, {Rossi}, {Bodo}, {Zanni}
  \& {Massaglia}}{{Capetti} et~al.}{2002}]{2002A&A...394...39C}
{Capetti} A.,  {Zamfir} S.,  {Rossi} P.,  {Bodo} G.,  {Zanni} C.,   {Massaglia}
  S.,  2002, \mn@doi [\aap] {10.1051/0004-6361:20021070}, \href
  {https://ui.adsabs.harvard.edu/abs/2002A&A...394...39C} {394, 39}

\bibitem[\protect\citeauthoryear{{CASA Team} et~al.,}{{CASA Team}
  et~al.}{2022}]{2022PASP..134k4501C}
{CASA Team} et~al., 2022, \mn@doi [\pasp] {10.1088/1538-3873/ac9642}, \href
  {https://ui.adsabs.harvard.edu/abs/2022PASP..134k4501C} {134, 114501}

\bibitem[\protect\citeauthoryear{{Chambers} et~al.,}{{Chambers}
  et~al.}{2016}]{2016arXiv161205560C}
{Chambers} K.~C.,  et~al., 2016, arXiv e-prints, \href
  {https://ui.adsabs.harvard.edu/abs/2016arXiv161205560C} {p. arXiv:1612.05560}

\bibitem[\protect\citeauthoryear{{Cheung}}{{Cheung}}{2007}]{2007AJ....133.2097C}
{Cheung} C.~C.,  2007, \mn@doi [\aj] {10.1086/513095}, \href
  {https://ui.adsabs.harvard.edu/abs/2007AJ....133.2097C} {133, 2097}

\bibitem[\protect\citeauthoryear{{Clark}}{{Clark}}{1980}]{1980A&A....89..377C}
{Clark} B.~G.,  1980, \aap, \href
  {https://ui.adsabs.harvard.edu/abs/1980A&A....89..377C} {89, 377}

\bibitem[\protect\citeauthoryear{{Comerford}, {Nevin}, {Stemo},
  {M{\"u}ller-S{\'a}nchez}, {Barrows}, {Cooper}  \& {Newman}}{{Comerford}
  et~al.}{2018}]{2018ApJ...867...66C}
{Comerford} J.~M.,  {Nevin} R.,  {Stemo} A.,  {M{\"u}ller-S{\'a}nchez} F.,
  {Barrows} R.~S.,  {Cooper} M.~C.,   {Newman} J.~A.,  2018, \mn@doi [\apj]
  {10.3847/1538-4357/aae2b4}, \href
  {https://ui.adsabs.harvard.edu/abs/2018ApJ...867...66C} {867, 66}

\bibitem[\protect\citeauthoryear{{Condon}, {Cotton}, {Greisen}, {Yin},
  {Perley}, {Taylor}  \& {Broderick}}{{Condon}
  et~al.}{1998}]{1998AJ....115.1693C}
{Condon} J.~J.,  {Cotton} W.~D.,  {Greisen} E.~W.,  {Yin} Q.~F.,  {Perley}
  R.~A.,  {Taylor} G.~B.,   {Broderick} J.~J.,  1998, \mn@doi [\aj]
  {10.1086/300337}, \href
  {https://ui.adsabs.harvard.edu/abs/1998AJ....115.1693C} {115, 1693}

\bibitem[\protect\citeauthoryear{{Cordey}}{{Cordey}}{1986}]{1986MNRAS.219..575C}
{Cordey} R.~A.,  1986, \mn@doi [\mnras] {10.1093/mnras/219.3.575}, \href
  {https://ui.adsabs.harvard.edu/abs/1986MNRAS.219..575C} {219, 575}

\bibitem[\protect\citeauthoryear{{Cotini}, {Ripamonti}, {Caccianiga}, {Colpi},
  {Della Ceca}, {Mapelli}, {Severgnini}  \& {Segreto}}{{Cotini}
  et~al.}{2013}]{2013MNRAS.431.2661C}
{Cotini} S.,  {Ripamonti} E.,  {Caccianiga} A.,  {Colpi} M.,  {Della Ceca} R.,
  {Mapelli} M.,  {Severgnini} P.,   {Segreto} A.,  2013, \mn@doi [\mnras]
  {10.1093/mnras/stt358}, \href
  {https://ui.adsabs.harvard.edu/abs/2013MNRAS.431.2661C} {431, 2661}

\bibitem[\protect\citeauthoryear{{Cotton}}{{Cotton}}{1999}]{1999ASPC..180..357C}
{Cotton} W.~D.,  1999, in {Taylor} G.~B.,  {Carilli} C.~L.,   {Perley} R.~A.,
  eds,  Astronomical Society of the Pacific Conference Series Vol. 180,
  Synthesis Imaging in Radio Astronomy II. p.~357

\bibitem[\protect\citeauthoryear{{Cotton} et~al.,}{{Cotton}
  et~al.}{2020}]{2020MNRAS.495.1271C}
{Cotton} W.~D.,  et~al., 2020, \mn@doi [\mnras] {10.1093/mnras/staa1240}, \href
  {https://ui.adsabs.harvard.edu/abs/2020MNRAS.495.1271C} {495, 1271}

\bibitem[\protect\citeauthoryear{{Croston}, {Hardcastle}, {Harris}, {Belsole},
  {Birkinshaw}  \& {Worrall}}{{Croston} et~al.}{2005}]{2005ApJ...626..733C}
{Croston} J.~H.,  {Hardcastle} M.~J.,  {Harris} D.~E.,  {Belsole} E.,
  {Birkinshaw} M.,   {Worrall} D.~M.,  2005, \mn@doi [\apj] {10.1086/430170},
  \href {https://ui.adsabs.harvard.edu/abs/2005ApJ...626..733C} {626, 733}

\bibitem[\protect\citeauthoryear{{de Gasperin} et~al.,}{{de Gasperin}
  et~al.}{2021}]{2021A&A...648A.104D}
{de Gasperin} F.,  et~al., 2021, \mn@doi [\aap] {10.1051/0004-6361/202140316},
  \href {https://ui.adsabs.harvard.edu/abs/2021A&A...648A.104D} {648, A104}

\bibitem[\protect\citeauthoryear{{Dennett-Thorpe}, {Scheuer}, {Laing},
  {Bridle}, {Pooley}  \& {Reich}}{{Dennett-Thorpe}
  et~al.}{2002}]{2002MNRAS.330..609D}
{Dennett-Thorpe} J.,  {Scheuer} P.~A.~G.,  {Laing} R.~A.,  {Bridle} A.~H.,
  {Pooley} G.~G.,   {Reich} W.,  2002, \mn@doi [\mnras]
  {10.1046/j.1365-8711.2002.05106.x}, \href
  {https://ui.adsabs.harvard.edu/abs/2002MNRAS.330..609D} {330, 609}

\bibitem[\protect\citeauthoryear{{Ekers}, {Fanti}, {Lari}  \& {Parma}}{{Ekers}
  et~al.}{1978}]{1978Natur.276..588E}
{Ekers} R.~D.,  {Fanti} R.,  {Lari} C.,   {Parma} P.,  1978, \mn@doi [\nat]
  {10.1038/276588a0}, \href
  {https://ui.adsabs.harvard.edu/abs/1978Natur.276..588E} {276, 588}

\bibitem[\protect\citeauthoryear{{Ellison}, {Viswanathan}, {Patton},
  {Bottrell}, {McConnachie}, {Gwyn}  \& {Cuillandre}}{{Ellison}
  et~al.}{2019}]{2019MNRAS.487.2491E}
{Ellison} S.~L.,  {Viswanathan} A.,  {Patton} D.~R.,  {Bottrell} C.,
  {McConnachie} A.~W.,  {Gwyn} S.,   {Cuillandre} J.-C.,  2019, \mn@doi
  [\mnras] {10.1093/mnras/stz1431}, \href
  {https://ui.adsabs.harvard.edu/abs/2019MNRAS.487.2491E} {487, 2491}

\bibitem[\protect\citeauthoryear{{Emonts}, {Morganti}, {Tadhunter},
  {Oosterloo}, {Holt}  \& {van der Hulst}}{{Emonts}
  et~al.}{2005}]{2005MNRAS.362..931E}
{Emonts} B.~H.~C.,  {Morganti} R.,  {Tadhunter} C.~N.,  {Oosterloo} T.~A.,
  {Holt} J.,   {van der Hulst} J.~M.,  2005, \mn@doi [\mnras]
  {10.1111/j.1365-2966.2005.09354.x}, \href
  {https://ui.adsabs.harvard.edu/abs/2005MNRAS.362..931E} {362, 931}

\bibitem[\protect\citeauthoryear{{Evans} et~al.,}{{Evans}
  et~al.}{2008}]{2008ApJ...675.1057E}
{Evans} D.~A.,  et~al., 2008, \mn@doi [\apj] {10.1086/527410}, \href
  {https://ui.adsabs.harvard.edu/abs/2008ApJ...675.1057E} {675, 1057}

\bibitem[\protect\citeauthoryear{{Fanaroff} \& {Riley}}{{Fanaroff} \&
  {Riley}}{1974}]{1974MNRAS.167P..31F}
{Fanaroff} B.~L.,  {Riley} J.~M.,  1974, \mn@doi [\mnras]
  {10.1093/mnras/167.1.31P}, \href
  {https://ui.adsabs.harvard.edu/abs/1974MNRAS.167P..31F} {167, 31P}

\bibitem[\protect\citeauthoryear{{Florido}, {Battaner}  \&
  {Sanchez-Saavedra}}{{Florido} et~al.}{1990}]{1990Ap&SS.164..131F}
{Florido} E.,  {Battaner} E.,   {Sanchez-Saavedra} M.~L.,  1990, \mn@doi
  [\apss] {10.1007/BF00653558}, \href
  {https://ui.adsabs.harvard.edu/abs/1990Ap&SS.164..131F} {164, 131}

\bibitem[\protect\citeauthoryear{{Gopal-Krishna}, {Biermann}, {Gergely}  \&
  {Wiita}}{{Gopal-Krishna} et~al.}{2012}]{2012RAA....12..127G}
{Gopal-Krishna} {Biermann} P.~L.,  {Gergely} L.~{\'A}.,   {Wiita} P.~J.,  2012,
  \mn@doi [Research in Astronomy and Astrophysics]
  {10.1088/1674-4527/12/2/002}, \href
  {https://ui.adsabs.harvard.edu/abs/2012RAA....12..127G} {12, 127}

\bibitem[\protect\citeauthoryear{{Greisen}}{{Greisen}}{2003}]{2003ASSL..285..109G}
{Greisen} E.~W.,  2003, in {Heck} A.,  ed.,  Astrophysics and Space Science
  Library Vol. 285, Information Handling in Astronomy - Historical Vistas.
  p.~109, \mn@doi{10.1007/0-306-48080-8_7}

\bibitem[\protect\citeauthoryear{{Hardcastle}, {Birkinshaw}, {Cameron},
  {Harris}, {Looney}  \& {Worrall}}{{Hardcastle}
  et~al.}{2002}]{2002ApJ...581..948H}
{Hardcastle} M.~J.,  {Birkinshaw} M.,  {Cameron} R.~A.,  {Harris} D.~E.,
  {Looney} L.~W.,   {Worrall} D.~M.,  2002, \mn@doi [\apj] {10.1086/344409},
  \href {https://ui.adsabs.harvard.edu/abs/2002ApJ...581..948H} {581, 948}

\bibitem[\protect\citeauthoryear{{Harris} \& {Grindlay}}{{Harris} \&
  {Grindlay}}{1979}]{1979MNRAS.188...25H}
{Harris} D.~E.,  {Grindlay} J.~E.,  1979, \mn@doi [\mnras]
  {10.1093/mnras/188.1.25}, \href
  {https://ui.adsabs.harvard.edu/abs/1979MNRAS.188...25H} {188, 25}

\bibitem[\protect\citeauthoryear{{Healey}, {Romani}, {Taylor}, {Sadler},
  {Ricci}, {Murphy}, {Ulvestad}  \& {Winn}}{{Healey}
  et~al.}{2007}]{2007ApJS..171...61H}
{Healey} S.~E.,  {Romani} R.~W.,  {Taylor} G.~B.,  {Sadler} E.~M.,  {Ricci} R.,
   {Murphy} T.,  {Ulvestad} J.~S.,   {Winn} J.~N.,  2007, \mn@doi [\apjs]
  {10.1086/513742}, \href
  {https://ui.adsabs.harvard.edu/abs/2007ApJS..171...61H} {171, 61}

\bibitem[\protect\citeauthoryear{{H{\"o}gbom}}{{H{\"o}gbom}}{1974}]{1974A&AS...15..417H}
{H{\"o}gbom} J.~A.,  1974, \aaps, \href
  {https://ui.adsabs.harvard.edu/abs/1974A&AS...15..417H} {15, 417}

\bibitem[\protect\citeauthoryear{{Horton}, {Krause}  \& {Hardcastle}}{{Horton}
  et~al.}{2020}]{2020MNRAS.499.5765H}
{Horton} M.~A.,  {Krause} M. G.~H.,   {Hardcastle} M.~J.,  2020, \mn@doi
  [\mnras] {10.1093/mnras/staa3020}, \href
  {https://ui.adsabs.harvard.edu/abs/2020MNRAS.499.5765H} {499, 5765}

\bibitem[\protect\citeauthoryear{{Hurley-Walker} et~al.,}{{Hurley-Walker}
  et~al.}{2015}]{2015MNRAS.447.2468H}
{Hurley-Walker} N.,  et~al., 2015, \mn@doi [\mnras] {10.1093/mnras/stu2570},
  \href {https://ui.adsabs.harvard.edu/abs/2015MNRAS.447.2468H} {447, 2468}

\bibitem[\protect\citeauthoryear{{Ineson}, {Croston}, {Hardcastle}  \&
  {Mingo}}{{Ineson} et~al.}{2017}]{2017MNRAS.467.1586I}
{Ineson} J.,  {Croston} J.~H.,  {Hardcastle} M.~J.,   {Mingo} B.,  2017,
  \mn@doi [\mnras] {10.1093/mnras/stx189}, \href
  {https://ui.adsabs.harvard.edu/abs/2017MNRAS.467.1586I} {467, 1586}

\bibitem[\protect\citeauthoryear{{Intema}}{{Intema}}{2014}]{2014ascl.soft08006I}
{Intema} H.~T.,  2014, {SPAM: Source Peeling and Atmospheric Modeling},
  Astrophysics Source Code Library, record ascl:1408.006 (\mn@eprint {ascl}
  {1408.006})

\bibitem[\protect\citeauthoryear{{Intema}, {van der Tol}, {Cotton}, {Cohen},
  {van Bemmel}  \& {R{\"o}ttgering}}{{Intema}
  et~al.}{2009}]{2009A&A...501.1185I}
{Intema} H.~T.,  {van der Tol} S.,  {Cotton} W.~D.,  {Cohen} A.~S.,  {van
  Bemmel} I.~M.,   {R{\"o}ttgering} H.~J.~A.,  2009, \mn@doi [\aap]
  {10.1051/0004-6361/200811094}, \href
  {https://ui.adsabs.harvard.edu/abs/2009A&A...501.1185I} {501, 1185}

\bibitem[\protect\citeauthoryear{{Jaffe} \& {Perola}}{{Jaffe} \&
  {Perola}}{1973}]{1973A&A....26..423J}
{Jaffe} W.~J.,  {Perola} G.~C.,  1973, \aap, \href
  {https://ui.adsabs.harvard.edu/abs/1973A&A....26..423J} {26, 423}

\bibitem[\protect\citeauthoryear{{Kaastra} \& {Roos}}{{Kaastra} \&
  {Roos}}{1992}]{1992A&A...254...96K}
{Kaastra} J.~S.,  {Roos} N.,  1992, \aap, \href
  {https://ui.adsabs.harvard.edu/abs/1992A&A...254...96K} {254, 96}

\bibitem[\protect\citeauthoryear{{Kataoka} \& {Stawarz}}{{Kataoka} \&
  {Stawarz}}{2005}]{2005ApJ...622..797K}
{Kataoka} J.,  {Stawarz} {\L}.,  2005, \mn@doi [\apj] {10.1086/428083}, \href
  {https://ui.adsabs.harvard.edu/abs/2005ApJ...622..797K} {622, 797}

\bibitem[\protect\citeauthoryear{{Kellermann}}{{Kellermann}}{1966}]{1966AuJPh..19..195K}
{Kellermann} K.~I.,  1966, \mn@doi [Australian Journal of Physics]
  {10.1071/PH660195}, \href
  {https://ui.adsabs.harvard.edu/abs/1966AuJPh..19..195K} {19, 195}

\bibitem[\protect\citeauthoryear{{Konar}, {Hardcastle}, {Croston}, {Jamrozy},
  {Hota}  \& {Das}}{{Konar} et~al.}{2019}]{2019MNRAS.486.3975K}
{Konar} C.,  {Hardcastle} M.~J.,  {Croston} J.~H.,  {Jamrozy} M.,  {Hota} A.,
  {Das} T.~K.,  2019, \mn@doi [\mnras] {10.1093/mnras/stz1089}, \href
  {https://ui.adsabs.harvard.edu/abs/2019MNRAS.486.3975K} {486, 3975}

\bibitem[\protect\citeauthoryear{{Kozie{\l}-Wierzbowska}, {Jamrozy}, {Zola},
  {Stachowski}  \& {Ku{\'z}micz}}{{Kozie{\l}-Wierzbowska}
  et~al.}{2012}]{2012MNRAS.422.1546K}
{Kozie{\l}-Wierzbowska} D.,  {Jamrozy} M.,  {Zola} S.,  {Stachowski} G.,
  {Ku{\'z}micz} A.,  2012, \mn@doi [\mnras] {10.1111/j.1365-2966.2012.20727.x},
  \href {https://ui.adsabs.harvard.edu/abs/2012MNRAS.422.1546K} {422, 1546}

\bibitem[\protect\citeauthoryear{{Ku{\'z}micz}, {Jamrozy},
  {Kozie{\l}-Wierzbowska}  \& {We{\.z}gowiec}}{{Ku{\'z}micz}
  et~al.}{2017}]{2017MNRAS.471.3806K}
{Ku{\'z}micz} A.,  {Jamrozy} M.,  {Kozie{\l}-Wierzbowska} D.,   {We{\.z}gowiec}
  M.,  2017, \mn@doi [\mnras] {10.1093/mnras/stx1830}, \href
  {https://ui.adsabs.harvard.edu/abs/2017MNRAS.471.3806K} {471, 3806}

\bibitem[\protect\citeauthoryear{{Lacy} et~al.,}{{Lacy}
  et~al.}{2020}]{2020PASP..132c5001L}
{Lacy} M.,  et~al., 2020, \mn@doi [\pasp] {10.1088/1538-3873/ab63eb}, \href
  {https://ui.adsabs.harvard.edu/abs/2020PASP..132c5001L} {132, 035001}

\bibitem[\protect\citeauthoryear{{Lal} \& {Rao}}{{Lal} \&
  {Rao}}{2005}]{2005MNRAS.356..232L}
{Lal} D.~V.,  {Rao} A.~P.,  2005, \mn@doi [\mnras]
  {10.1111/j.1365-2966.2004.08442.x}, \href
  {https://ui.adsabs.harvard.edu/abs/2005MNRAS.356..232L} {356, 232}

\bibitem[\protect\citeauthoryear{{Leahy} \& {Parma}}{{Leahy} \&
  {Parma}}{1992}]{1992ersf.meet..307L}
{Leahy} J.~P.,  {Parma} P.,  1992, in {Roland} J.,  {Sol} H.,   {Pelletier} G.,
   eds, Extragalactic Radio Sources. From Beams to Jets. pp 307--308

\bibitem[\protect\citeauthoryear{{Leahy} \& {Williams}}{{Leahy} \&
  {Williams}}{1984}]{1984MNRAS.210..929L}
{Leahy} J.~P.,  {Williams} A.~G.,  1984, \mn@doi [\mnras]
  {10.1093/mnras/210.4.929}, \href
  {https://ui.adsabs.harvard.edu/abs/1984MNRAS.210..929L} {210, 929}

\bibitem[\protect\citeauthoryear{{Ledlow}, {Owen}, {Yun}  \& {Hill}}{{Ledlow}
  et~al.}{2001}]{2001ApJ...552..120L}
{Ledlow} M.~J.,  {Owen} F.~N.,  {Yun} M.~S.,   {Hill} J.~M.,  2001, \mn@doi
  [\apj] {10.1086/320458}, \href
  {https://ui.adsabs.harvard.edu/abs/2001ApJ...552..120L} {552, 120}

\bibitem[\protect\citeauthoryear{{Lu}}{{Lu}}{1990}]{1990A&A...229..424L}
{Lu} J.~F.,  1990, \aap, \href
  {https://ui.adsabs.harvard.edu/abs/1990A&A...229..424L} {229, 424}

\bibitem[\protect\citeauthoryear{{Machalski}, {Jamrozy}  \&
  {Saikia}}{{Machalski} et~al.}{2009}]{2009MNRAS.395..812M}
{Machalski} J.,  {Jamrozy} M.,   {Saikia} D.~J.,  2009, \mn@doi [\mnras]
  {10.1111/j.1365-2966.2009.14516.x}, \href
  {https://ui.adsabs.harvard.edu/abs/2009MNRAS.395..812M} {395, 812}

\bibitem[\protect\citeauthoryear{{Machalski}, {Kozie{\l}-Wierzbowska}  \&
  {Goyal}}{{Machalski} et~al.}{2021}]{2021ApJS..255...22M}
{Machalski} J.,  {Kozie{\l}-Wierzbowska} D.,   {Goyal} A.,  2021, \mn@doi
  [\apjs] {10.3847/1538-4365/ac08a0}, \href
  {https://ui.adsabs.harvard.edu/abs/2021ApJS..255...22M} {255, 22}

\bibitem[\protect\citeauthoryear{{Mahatma} et~al.,}{{Mahatma}
  et~al.}{2019}]{2019A&A...622A..13M}
{Mahatma} V.~H.,  et~al., 2019, \mn@doi [\aap] {10.1051/0004-6361/201833973},
  \href {https://ui.adsabs.harvard.edu/abs/2019A&A...622A..13M} {622, A13}

\bibitem[\protect\citeauthoryear{{Merritt} \& {Ekers}}{{Merritt} \&
  {Ekers}}{2002}]{2002Sci...297.1310M}
{Merritt} D.,  {Ekers} R.~D.,  2002, \mn@doi [Science]
  {10.1126/science.1074688}, \href
  {https://ui.adsabs.harvard.edu/abs/2002Sci...297.1310M} {297, 1310}

\bibitem[\protect\citeauthoryear{{Merritt} \& {Milosavljevi{\'c}}}{{Merritt} \&
  {Milosavljevi{\'c}}}{2005}]{2005LRR.....8....8M}
{Merritt} D.,  {Milosavljevi{\'c}} M.,  2005, \mn@doi [Living Reviews in
  Relativity] {10.12942/lrr-2005-8}, \href
  {https://ui.adsabs.harvard.edu/abs/2005LRR.....8....8M} {8, 8}

\bibitem[\protect\citeauthoryear{{Mezcua}, {Chavushyan}, {Lobanov}  \&
  {Le{\'o}n-Tavares}}{{Mezcua} et~al.}{2012}]{2012A&A...544A..36M}
{Mezcua} M.,  {Chavushyan} V.~H.,  {Lobanov} A.~P.,   {Le{\'o}n-Tavares} J.,
  2012, \mn@doi [\aap] {10.1051/0004-6361/201117724}, \href
  {https://ui.adsabs.harvard.edu/abs/2012A&A...544A..36M} {544, A36}

\bibitem[\protect\citeauthoryear{{Migliori}, {Grandi}, {Palumbo}, {Brunetti}
  \& {Stanghellini}}{{Migliori} et~al.}{2007}]{2007ApJ...668..203M}
{Migliori} G.,  {Grandi} P.,  {Palumbo} G. G.~C.,  {Brunetti} G.,
  {Stanghellini} C.,  2007, \mn@doi [\apj] {10.1086/520870}, \href
  {https://ui.adsabs.harvard.edu/abs/2007ApJ...668..203M} {668, 203}

\bibitem[\protect\citeauthoryear{{Murgia}}{{Murgia}}{1996}]{1996PhDT........92M}
{Murgia} M.,  1996, PhD thesis, -

\bibitem[\protect\citeauthoryear{{Murgia}}{{Murgia}}{2003}]{2003PASA...20...19M}
{Murgia} M.,  2003, \mn@doi [\pasa] {10.1071/AS02033}, \href
  {https://ui.adsabs.harvard.edu/abs/2003PASA...20...19M} {20, 19}

\bibitem[\protect\citeauthoryear{{Murgia} et~al.,}{{Murgia}
  et~al.}{2011}]{2011A&A...526A.148M}
{Murgia} M.,  et~al., 2011, \mn@doi [\aap] {10.1051/0004-6361/201015302}, \href
  {https://ui.adsabs.harvard.edu/abs/2011A&A...526A.148M} {526, A148}

\bibitem[\protect\citeauthoryear{{Myers} et~al.,}{{Myers}
  et~al.}{2003}]{2003MNRAS.341....1M}
{Myers} S.~T.,  et~al., 2003, \mn@doi [\mnras]
  {10.1046/j.1365-8711.2003.06256.x}, \href
  {https://ui.adsabs.harvard.edu/abs/2003MNRAS.341....1M} {341, 1}

\bibitem[\protect\citeauthoryear{{O'Dea}}{{O'Dea}}{1998}]{1998PASP..110..493O}
{O'Dea} C.~P.,  1998, \mn@doi [\pasp] {10.1086/316162}, \href
  {https://ui.adsabs.harvard.edu/abs/1998PASP..110..493O} {110, 493}

\bibitem[\protect\citeauthoryear{{Pacholczyk}}{{Pacholczyk}}{1970}]{1970ranp.book.....P}
{Pacholczyk} A.~G.,  1970, {Radio astrophysics. Nonthermal processes in
  galactic and extragalactic sources}

\bibitem[\protect\citeauthoryear{{Parma}, {Ekers}  \& {Fanti}}{{Parma}
  et~al.}{1985}]{1985A&AS...59..511P}
{Parma} P.,  {Ekers} R.~D.,   {Fanti} R.,  1985, \aaps, \href
  {https://ui.adsabs.harvard.edu/abs/1985A&AS...59..511P} {59, 511}

\bibitem[\protect\citeauthoryear{{Parma}, {Murgia}, {Morganti}, {Capetti}, {de
  Ruiter}  \& {Fanti}}{{Parma} et~al.}{1999}]{1999A&A...344....7P}
{Parma} P.,  {Murgia} M.,  {Morganti} R.,  {Capetti} A.,  {de Ruiter} H.~R.,
  {Fanti} R.,  1999, \aap, \href
  {https://ui.adsabs.harvard.edu/abs/1999A&A...344....7P} {344, 7}

\bibitem[\protect\citeauthoryear{{Parma}, {Murgia}, {de Ruiter}, {Fanti},
  {Mack}  \& {Govoni}}{{Parma} et~al.}{2007}]{2007A&A...470..875P}
{Parma} P.,  {Murgia} M.,  {de Ruiter} H.~R.,  {Fanti} R.,  {Mack} K.~H.,
  {Govoni} F.,  2007, \mn@doi [\aap] {10.1051/0004-6361:20077592}, \href
  {https://ui.adsabs.harvard.edu/abs/2007A&A...470..875P} {470, 875}

\bibitem[\protect\citeauthoryear{{Pringle}}{{Pringle}}{1997}]{1997MNRAS.292..136P}
{Pringle} J.~E.,  1997, \mn@doi [\mnras] {10.1093/mnras/292.1.136}, \href
  {https://ui.adsabs.harvard.edu/abs/1997MNRAS.292..136P} {292, 136}

\bibitem[\protect\citeauthoryear{{Romero}, {Chajet}, {Abraham}  \&
  {Fan}}{{Romero} et~al.}{2000}]{2000A&A...360...57R}
{Romero} G.~E.,  {Chajet} L.,  {Abraham} Z.,   {Fan} J.~H.,  2000, \aap, \href
  {https://ui.adsabs.harvard.edu/abs/2000A&A...360...57R} {360, 57}

\bibitem[\protect\citeauthoryear{{Roos}, {Kaastra}  \& {Hummel}}{{Roos}
  et~al.}{1993}]{1993ApJ...409..130R}
{Roos} N.,  {Kaastra} J.~S.,   {Hummel} C.~A.,  1993, \mn@doi [\apj]
  {10.1086/172647}, \href
  {https://ui.adsabs.harvard.edu/abs/1993ApJ...409..130R} {409, 130}

\bibitem[\protect\citeauthoryear{{Rottmann}}{{Rottmann}}{2002}]{2002PhDT.......178R}
{Rottmann} H.,  2002, PhD thesis, University of Bonn

\bibitem[\protect\citeauthoryear{{Saikia} \& {Jamrozy}}{{Saikia} \&
  {Jamrozy}}{2009}]{2009BASI...37...63S}
{Saikia} D.~J.,  {Jamrozy} M.,  2009, Bulletin of the Astronomical Society of
  India, \href {https://ui.adsabs.harvard.edu/abs/2009BASI...37...63S} {37, 63}

\bibitem[\protect\citeauthoryear{{Saripalli} \& {Roberts}}{{Saripalli} \&
  {Roberts}}{2018}]{2018ApJ...852...48S}
{Saripalli} L.,  {Roberts} D.~H.,  2018, \mn@doi [\apj]
  {10.3847/1538-4357/aa9c4b}, \href
  {https://ui.adsabs.harvard.edu/abs/2018ApJ...852...48S} {852, 48}

\bibitem[\protect\citeauthoryear{{Schoenmakers}, {de Bruyn}, {R{\"o}ttgering},
  {van der Laan}  \& {Kaiser}}{{Schoenmakers}
  et~al.}{2000}]{2000MNRAS.315..371S}
{Schoenmakers} A.~P.,  {de Bruyn} A.~G.,  {R{\"o}ttgering} H.~J.~A.,  {van der
  Laan} H.,   {Kaiser} C.~R.,  2000, \mn@doi [\mnras]
  {10.1046/j.1365-8711.2000.03430.x}, \href
  {https://ui.adsabs.harvard.edu/abs/2000MNRAS.315..371S} {315, 371}

\bibitem[\protect\citeauthoryear{{Schwab}}{{Schwab}}{1984}]{1984AJ.....89.1076S}
{Schwab} F.~R.,  1984, \mn@doi [\aj] {10.1086/113605}, \href
  {https://ui.adsabs.harvard.edu/abs/1984AJ.....89.1076S} {89, 1076}

\bibitem[\protect\citeauthoryear{{Shimwell} et~al.,}{{Shimwell}
  et~al.}{2022}]{2022A&A...659A...1S}
{Shimwell} T.~W.,  et~al., 2022, \mn@doi [\aap] {10.1051/0004-6361/202142484},
  \href {https://ui.adsabs.harvard.edu/abs/2022A&A...659A...1S} {659, A1}

\bibitem[\protect\citeauthoryear{{Shulevski} et~al.,}{{Shulevski}
  et~al.}{2019}]{2019A&A...628A..69S}
{Shulevski} A.,  et~al., 2019, \mn@doi [\aap] {10.1051/0004-6361/201935586},
  \href {https://ui.adsabs.harvard.edu/abs/2019A&A...628A..69S} {628, A69}

\bibitem[\protect\citeauthoryear{{Singh}, {Ishwara-Chandra}, {Sievers},
  {Wadadekar}, {Hilton}  \& {Beelen}}{{Singh}
  et~al.}{2015}]{2015MNRAS.454.1556S}
{Singh} V.,  {Ishwara-Chandra} C.~H.,  {Sievers} J.,  {Wadadekar} Y.,  {Hilton}
  M.,   {Beelen} A.,  2015, \mn@doi [\mnras] {10.1093/mnras/stv2071}, \href
  {https://ui.adsabs.harvard.edu/abs/2015MNRAS.454.1556S} {454, 1556}

\bibitem[\protect\citeauthoryear{{Skipper} \& {Browne}}{{Skipper} \&
  {Browne}}{2018}]{2018MNRAS.475.5179S}
{Skipper} C.~J.,  {Browne} I. W.~A.,  2018, \mn@doi [\mnras]
  {10.1093/mnras/sty114}, \href
  {https://ui.adsabs.harvard.edu/abs/2018MNRAS.475.5179S} {475, 5179}

\bibitem[\protect\citeauthoryear{Springmann \& Coll.}{Springmann \&
  Coll.}{2006}]{Springmann2006HOSTGO}
Springmann A.,  Coll. W.,  2006, The Journal of Undergraduate Research, 7

\bibitem[\protect\citeauthoryear{{Swarup}, {Ananthakrishnan}, {Kapahi}, {Rao},
  {Subrahmanya}  \& {Kulkarni}}{{Swarup} et~al.}{1991}]{1991CSci...60...95S}
{Swarup} G.,  {Ananthakrishnan} S.,  {Kapahi} V.~K.,  {Rao} A.~P.,
  {Subrahmanya} C.~R.,   {Kulkarni} V.~K.,  1991, Current Science, \href
  {https://ui.adsabs.harvard.edu/abs/1991CSci...60...95S} {60, 95}

\bibitem[\protect\citeauthoryear{{Tremblay}, {Taylor}, {Ortiz}, {Tremblay},
  {Helmboldt}  \& {Romani}}{{Tremblay} et~al.}{2016}]{2016MNRAS.459..820T}
{Tremblay} S.~E.,  {Taylor} G.~B.,  {Ortiz} A.~A.,  {Tremblay} C.~D.,
  {Helmboldt} J.~F.,   {Romani} R.~W.,  2016, \mn@doi [\mnras]
  {10.1093/mnras/stw592}, \href
  {https://ui.adsabs.harvard.edu/abs/2016MNRAS.459..820T} {459, 820}

\bibitem[\protect\citeauthoryear{{Turner}, {Shabala}  \& {Krause}}{{Turner}
  et~al.}{2018}]{2018MNRAS.474.3361T}
{Turner} R.~J.,  {Shabala} S.~S.,   {Krause} M. G.~H.,  2018, \mn@doi [\mnras]
  {10.1093/mnras/stx2947}, \href
  {https://ui.adsabs.harvard.edu/abs/2018MNRAS.474.3361T} {474, 3361}

\bibitem[\protect\citeauthoryear{{van Breugel}, {Balick}, {Heckman}, {Miley}
  \& {Helfand}}{{van Breugel} et~al.}{1983}]{1983AJ.....88...40V}
{van Breugel} W.,  {Balick} B.,  {Heckman} T.,  {Miley} G.,   {Helfand} D.,
  1983, \mn@doi [\aj] {10.1086/113285}, \href
  {https://ui.adsabs.harvard.edu/abs/1983AJ.....88...40V} {88, 40}

\bibitem[\protect\citeauthoryear{{van Haarlem} et~al.,}{{van Haarlem}
  et~al.}{2013}]{2013A&A...556A...2V}
{van Haarlem} M.~P.,  et~al., 2013, \mn@doi [\aap]
  {10.1051/0004-6361/201220873}, \href
  {https://ui.adsabs.harvard.edu/abs/2013A&A...556A...2V} {556, A2}

\bibitem[\protect\citeauthoryear{{Worrall}, {Birkinshaw}  \&
  {Cameron}}{{Worrall} et~al.}{1995}]{1995ApJ...449...93W}
{Worrall} D.~M.,  {Birkinshaw} M.,   {Cameron} R.~A.,  1995, \mn@doi [\apj]
  {10.1086/176035}, \href
  {https://ui.adsabs.harvard.edu/abs/1995ApJ...449...93W} {449, 93}

\bibitem[\protect\citeauthoryear{{Yang} et~al.,}{{Yang}
  et~al.}{2019}]{2019ApJS..245...17Y}
{Yang} X.,  et~al., 2019, \mn@doi [\apjs] {10.3847/1538-4365/ab4811}, \href
  {https://ui.adsabs.harvard.edu/abs/2019ApJS..245...17Y} {245, 17}

\bibitem[\protect\citeauthoryear{{Yang} et~al.,}{{Yang}
  et~al.}{2022}]{2022ApJ...933...98Y}
{Yang} X.,  et~al., 2022, \mn@doi [\apj] {10.3847/1538-4357/ac71aa}, \href
  {https://ui.adsabs.harvard.edu/abs/2022ApJ...933...98Y} {933, 98}

\bibitem[\protect\citeauthoryear{{Zou} et~al.,}{{Zou}
  et~al.}{2019}]{2019ApJS..245....4Z}
{Zou} H.,  et~al., 2019, \mn@doi [\apjs] {10.3847/1538-4365/ab48e8}, \href
  {https://ui.adsabs.harvard.edu/abs/2019ApJS..245....4Z} {245, 4}

\makeatother
\end{thebibliography}








\bsp	
\label{lastpage}

\end{document}